\documentclass[10pt,letterpaper]{article}

\usepackage{extsizes}
\usepackage[super,sort&compress,comma]{natbib} 
\usepackage[version=3]{mhchem}
\usepackage{balance}
\usepackage{mathptmx}
\usepackage{subfig}
\usepackage{sectsty}
\usepackage{graphicx} 
\usepackage{lastpage}
\usepackage{amssymb}
\usepackage[format=plain,justification=justified,singlelinecheck=false,font={stretch=1.125,small,sf},labelfont=bf,labelsep=space]{caption}
\usepackage{float}
\usepackage{fancyhdr}
\usepackage{fnpos}
\usepackage[english]{babel}
\addto{\captionsenglish}{%
  
}
\usepackage{array}
\usepackage{droidsans}
\usepackage{charter}
\usepackage[T1]{fontenc}
\usepackage[usenames,dvipsnames]{xcolor}
\usepackage{setspace}
\usepackage[compact]{titlesec}
\usepackage{bm} 

\usepackage{epstopdf}

\usepackage[top=1in, bottom=1in, left=1in, right=1in]{geometry}

\usepackage[utf8]{inputenc}

\usepackage{amsmath}
\usepackage{nameref,hyperref}

\usepackage[right]{lineno}

\usepackage{microtype}
\DisableLigatures[f]{encoding = *, family = * }

\usepackage[aboveskip=1pt,labelfont=bf,labelsep=period,singlelinecheck=off]{caption}

\makeatletter
\renewcommand{\@biblabel}[1]{\quad#1.}
\makeatother

\usepackage{lastpage,fancyhdr}
\usepackage{color}
\definecolor{Gray}{gray}{.25}
\definecolor{cream}{RGB}{222,217,201}

\usepackage{graphicx}
\usepackage{epstopdf}

\usepackage{sidecap}
\usepackage{wrapfig}

\begin{document}

\pagestyle{fancy}
\thispagestyle{plain}
\fancypagestyle{plain}{
\renewcommand{\headrulewidth}{0pt}
}

\makeFNbottom
\makeatletter
\renewcommand\LARGE{\@setfontsize\LARGE{15pt}{17}}
\renewcommand\Large{\@setfontsize\Large{12pt}{14}}
\renewcommand\large{\@setfontsize\large{10pt}{12}}
\renewcommand\footnotesize{\@setfontsize\footnotesize{7pt}{10}}
\makeatother

\renewcommand{\thefootnote}{\fnsymbol{footnote}}
\renewcommand\footnoterule{\vspace*{1pt}%
\color{cream}\hrule width 3.5in height 0.4pt \color{black}\vspace*{5pt}} 
\setcounter{secnumdepth}{5}

\makeatletter 
\renewcommand\@biblabel[1]{#1}            
\renewcommand\@makefntext[1]%
{\noindent\makebox[0pt][r]{\@thefnmark\,}#1}
\makeatother 
\renewcommand{\figurename}{\small{Fig.}~}
\sectionfont{\sffamily\Large}
\subsectionfont{\normalsize}
\subsubsectionfont{\bf}
\setstretch{1.125} 
\setlength{\skip\footins}{0.8cm}
\setlength{\footnotesep}{0.25cm}
\setlength{\jot}{10pt}
\titlespacing*{\section}{0pt}{4pt}{4pt}
\titlespacing*{\subsection}{0pt}{15pt}{1pt}


\makeatletter 
\newlength{\figrulesep} 
\setlength{\figrulesep}{0.5\textfloatsep} 

\newcommand{\topfigrule}{\vspace*{-1pt}%
\noindent{\color{cream}\rule[-\figrulesep]{\columnwidth}{1.5pt}} }

\newcommand{\botfigrule}{\vspace*{-2pt}%
\noindent{\color{cream}\rule[\figrulesep]{\columnwidth}{1.5pt}} }

\newcommand{\dblfigrule}{\vspace*{-1pt}%
\noindent{\color{cream}\rule[-\figrulesep]{\textwidth}{1.5pt}} }

\makeatother


 \noindent\LARGE{\textbf{Multiparticle Collision Dynamics Simulations of the Flagellar Apparatus in \textit{Chlamydomonas reinhardtii}}} \\

  \noindent\large{Sai Venkata Ramana Ambadipudi$^\text{a}$, Albert Bae$^\text{b}$, Azam Gholami$^{\text{*a}}$}\\

\section*{Abstract}
 Using multiparticle collision dynamics simulations, we investigate the swimming dynamics, orientational behavior, and hydrodynamic interactions of a model swimmer designed to mimic the isolated flagellar apparatus (\(FA\)) of \textit{Chlamydomonas reinhardtii}. We represent the \(FA\) as a chain of monomers connected by elastic springs, with two traveling waves originating at its center and propagating in opposite directions along the chain.  Our simulations show that an \(FA\) whose beat pattern has non-zero mean curvature sustains ballistic motion for several hundred beats before transitioning to a diffusion‐dominated regime via rotational diffusion. In contrast, a flagellar apparatus with zero mean curvature \(FA_{0}\)—generates mirror‐symmetric deformations and fails to achieve net propulsion.  Both the active \(FA\) and \(FA_{0}\) exhibit orientational autocorrelation functions that decay exponentially—matching those of their inactive counterparts—indicating that active beating does not influence the \(FA\)’s rotational diffusion. Driving the two flagellar arms at different frequencies reproduces the epitrochoid-like trajectory observed experimentally. Finally, hydrodynamic interactions between two \(FA\)s give rise to co‐moving bound pairs in either parallel or antiparallel configurations, with their stability governed by the phase difference of the curvature waves. Together, our results establish a versatile model microswimmer with tunable dynamics—offering a blueprint for the rational design of artificial, flagella‐driven microswimmers. \\


\renewcommand*\rmdefault{bch}\normalfont\upshape
\rmfamily
\section*{}
\vspace{-1cm}


	\footnotetext{$^\text{a}$New York University Abu Dhabi, Abu Dhabi, United Arab Emirates. $^\text{b}$ Department of Physics, Lewis \& Clark College, Portland, Oregon, USA. \\E-mail: azam.gholami@nyu.edu }
\footnotetext{\dag~Electronic Supplementary Information (ESI) available: See DOI: 00.0000/00000000.}



\section{Introduction}
Cilia and flagella are slender, whip-like appendages which protrude from the surface of many eukaryotic cells~\cite{gibbons1981cilia,gilpin2020multiscale} and generate propulsion through a characteristic beating motion, exploiting anisotropic viscous drag in the low Reynolds number regime~\cite{purcell2014life,blake2001fluid,lauga2009hydrodynamics,elgeti2015physics}. Understanding the hydrodynamics of flagella is crucial, as they play a fundamental role in the motility of various microorganisms, including bacteria and sperm cells~\cite{elgeti2010hydrodynamics,taketoshi2020elasto}. This knowledge is essential not only for studying biological systems such as sperm motility and bacterial swarming but also for applications in designing artificial microswimmers~\cite{liu2024propulsion,dreyfus2005microscopic,tsang2020roads,ahmad2022bio,bae2023flagellum}.

The dynamics of a single flagellum have been extensively studied through experiments~\cite{gray1955propulsion,fauci2006biofluidmechanics,hilfinger2009nonlinear,sartori2016curvature} and simulations~\cite{camalet1999self,brokaw2001simulating}, leading to significant advancements in understanding flagellar propulsion. Theoretical frameworks have successfully explained the mechanics of flagellar motion~\cite{elgeti2015physics,hilfinger2009nonlinear,riedel2007molecular}, offering insights into the underlying physics. Additionally, studies on collective behavior~\cite{vilfan2006hydrodynamic,hickey2023nonreciprocal,cheng2024near,hickey2021ciliary}, such as those by Yang \emph{et al.}
~\cite{yang2008cooperation,yang2010swarm}, have demonstrated that flagella or sperm cells moving in the same direction experience hydrodynamic attraction, synchronize their beating, and form clusters due to fluid-mediated interactions. These findings highlight the crucial role of hydrodynamic interactions in collective motility.

Despite this progress, the dynamics of flagella that are physically connected remain poorly understood. The coupling between flagella, combined with their hydrodynamic interactions, adds significant complexity to these systems. A key question—relevant both to biological motility and artificial microswimmer design—is whether propulsion efficiency increases with the number of flagella attached to a basal body~\cite{singleton2011micro,ye2013rotating,lauga2007floppy}. Additionally, studies have explored the optimal placement of flagella on the basal body~\cite{hu2024multiflagellate}. Nature provides examples of microswimmers with multiple flagella, such as \textit{Chlamydomonas reinhardtii}, \textit{E. coli}, \textit{Salmonella typhimurium}, and \textit{Rhizobium lupini}~\cite{berg2004coli,brennen1977fluid}. Whether these microorganisms evolved multiple flagella to optimize hydrodynamic interactions remains an open question~\cite{omori2020swimming,osterman2011finding}. Furthermore, the role of hydrodynamics in synchronizing flagellar beating when connected to a basal body is yet to be fully understood~\cite{goldstein2015green,wan2016coordinated,kendelbacher2013synchronization,wollin2011metachronal,liao2021energetics}.

\textit{Chlamydomonas reinhardtii}, a single-celled green alga with two flagella protruding from its cell body, serves as an excellent model system for studying the hydrodynamic interactions of physically connected flagella~\cite{harris2001chlamydomonas,drescher2010direct}. The two flagella of \textit{C. reinhardtii} typically beat in synchrony for extended periods before transitioning to an asynchronous state during re-orientation, after which they regain synchronization~\cite{polin2009chlamydomonas,goldstein2009noise,leptos2013antiphase}. While some experiments highlight the significance of hydrodynamic interactions in synchronization~\cite{uchida2011generic,golestanian2011hydrodynamic,vilfan2006hydrodynamic}, other experiments with \textit{C. reinhardtii}  also support the crucial role of mechanical coupling through basal bodies~\cite{quaranta2015hydrodynamics,ringo1967flagellar,melkonian1991development,geyer2013cell,friedrich2012flagellar,wan2016coordinated,wan2014rhythmicity}. The basal body is composed of elastic fibers with a microtubule-based structure exhibiting periodic striation patterns~\cite{ringo1967flagellar}. In \textit{C. reinhardtii}, these periodic striations are approximately 80 nm apart and have been shown to respond dynamically to chemical stimuli such as calcium ions, indicating the contractile nature of the fibers~\cite{sleigh1979contractility}.

Experiments on the isolated flagellar apparatus (\(FA\)s) of a wall-less mutant of \textit{C. reinhardtii}, conducted by Hyams and Borisy~\cite{hyams1975flagellar}, demonstrated that both flagella can sustain their characteristic beating patterns even in the absence of the cell body and cytoplasm. To facilitate microscopy, Hyams and Borisy primarily examined \(FA\)s anchored to substrate debris, observing that over \(70\%\) exhibited synchronous beating, while the remainder beat asynchronously. They also documented transient switches between synchronous and asynchronous states. More recently, Pozveh \emph{et al.}~\cite{pozveh2021resistive} investigated freely swimming \(FA\)s and found that when the frequency difference between the two flagella was substantial (\(10-41\%\) of the mean), neither mechanical coupling via the basal body nor hydrodynamic interactions were sufficient to achieve synchronization. Despite these insights, experiments with isolated \(FA\)s remain challenging due to their low yield, with successful isolation and reactivation of \(FA\)s occurring in only a small percentage of attempts. Furthermore, most isolated apparatuses exhibit frequency differences exceeding 15\%, making large-scale statistical analysis difficult. Given these limitations, developing a reliable computational model of the \(FA\)s is essential to complement experimental findings and enable systematic investigations.
\begin{figure*}[t!]
	\centering
	\includegraphics[width=1.0\columnwidth]{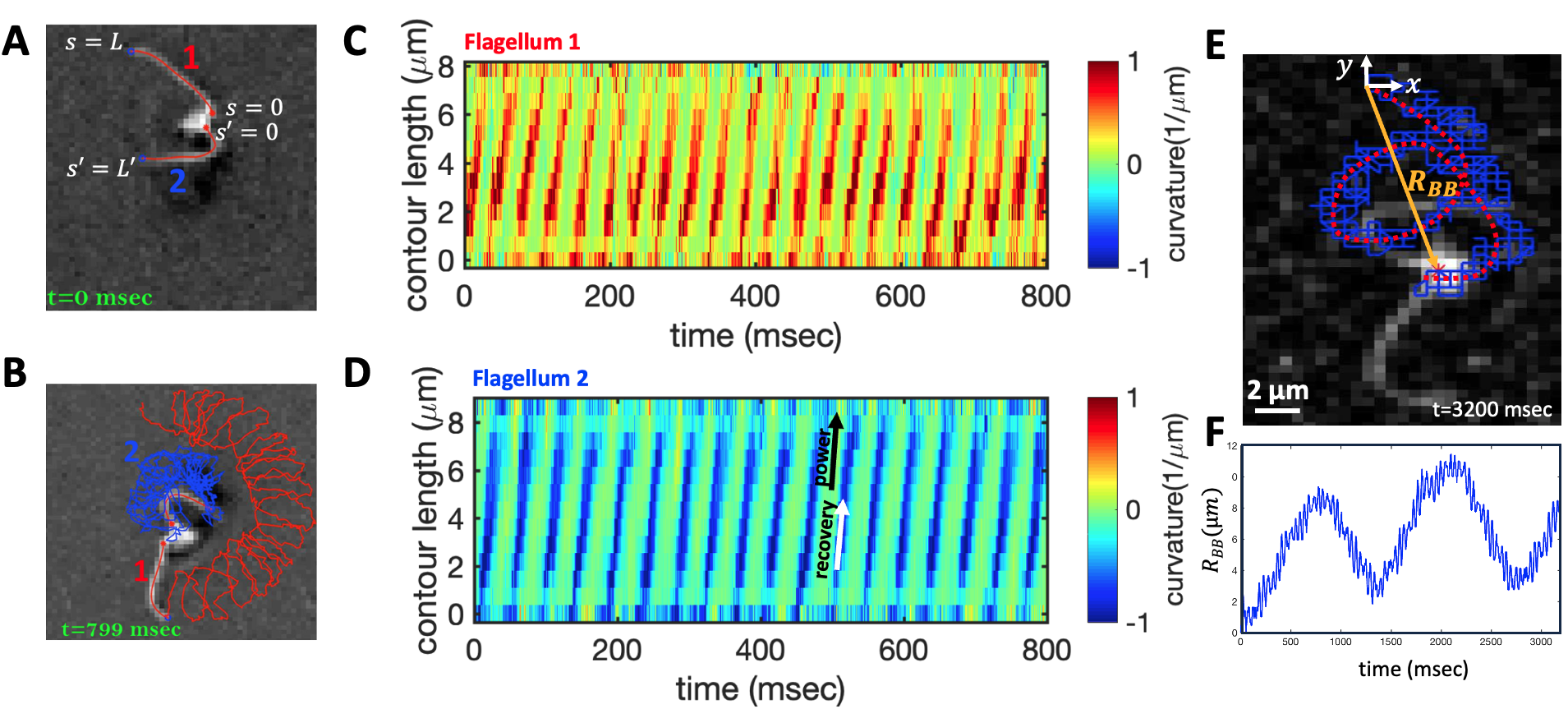}
	\includegraphics[width=1.0\columnwidth]{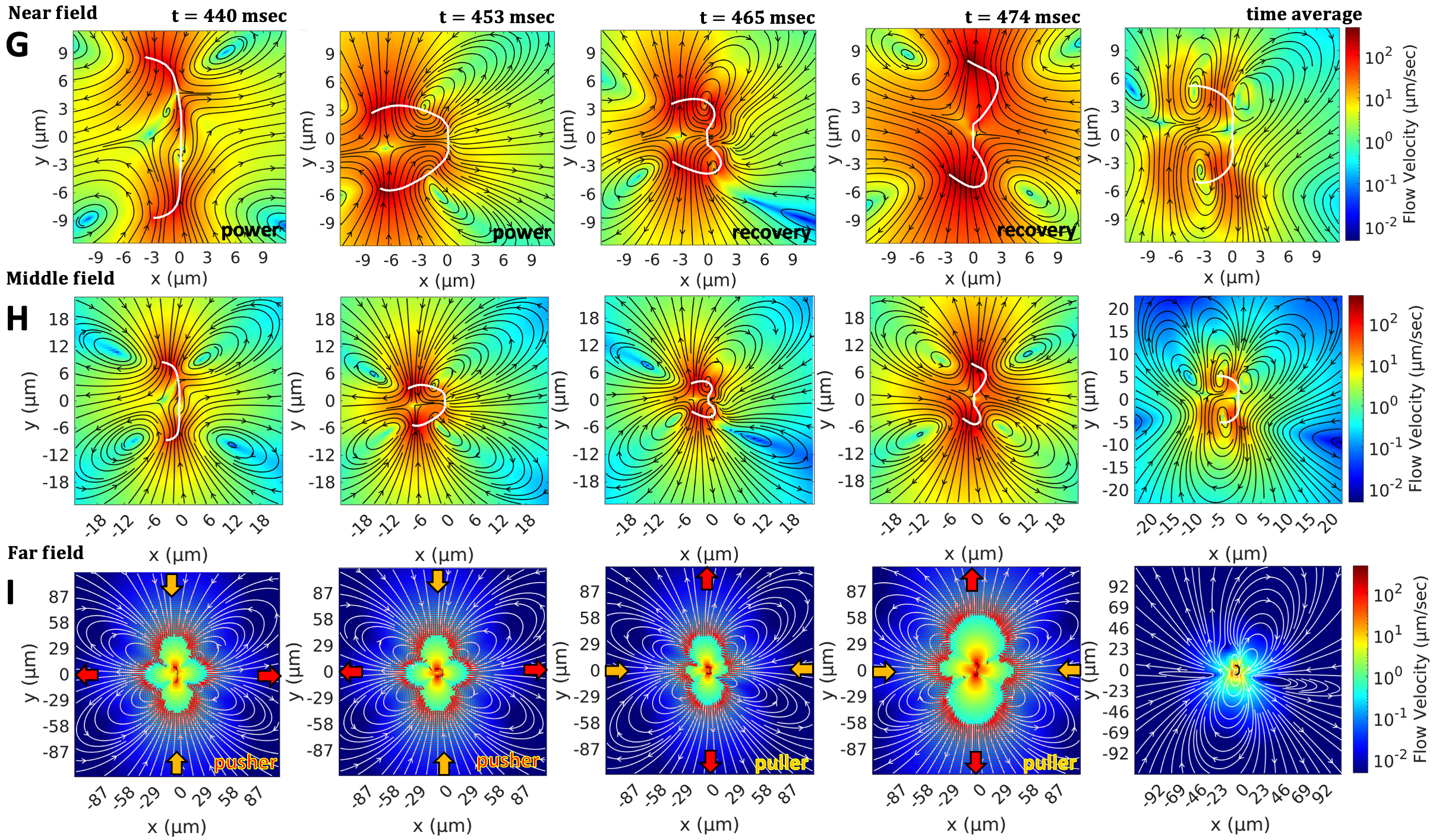}
	\caption{ A-B) Representative snapshots showing an isolated flagellar apparatus swimming in a water-like fluid supplemented with 2mM ATP (\textcolor{black}{see Video 1}). The red and blue trajectories in panel B trace the distal ends of flagella 1 and 2, respectively, as the flagellar apparatus swims. The first flagellum beats at the frequency of 22.50 Hz, while the second one beats faster at frequency of 26.25 Hz. C-D) Propagation of curvature waves along the contour length of both flagella, highlighting the characteristic power and recovery strokes. For each flagellum curvature waves initiate at the basal end at ($s,s'=0$) and propagate toward the distal tip ($s,s'=L,L'$). E-F) Over time, the geometric center of the basal body traces an epitrochoid-like trajectory. \textcolor{black}{G-I) The flow field is calculated throughout the 3D domain using experimental \(FA\) shapes swimming near a substrate (approximately at a height of 10~\(\mu\)m), based on the method of images for the regularized Stokeslet developed by Ainley \emph{et al.}~\cite{ainley2008method} (refer to Appendix~\ref{Stokslets}). Analysis of the far‐field flow shows that the \(FA\) alternates between puller and pusher modes, with an average behavior resembling that of a nearly neutral swimmer.
	} }
	\label{fig:FA_EXP}
\end{figure*}

In this study, we build upon the framework introduced by Yang \emph{et al.}\cite{yang2008cooperation} to develop a model for the \(FA\)s and employ multiparticle collision dynamics (MPC) simulations to examine their swimming dynamics. The paper is structured as follows: \textcolor{black}{Section~\ref{EXP} reviews the experimental observations that motivated this study.  Section~\ref{Stokselet} presents our new Stokeslet‐based flow‐field computations, which we compare qualitatively with the MPC results}. Section~\ref{MPC} presents the details of our modeling approach and the MPC methodology. In Section~\ref{FA}, we analyze the flow-field, the swimming dynamics and orientational behavior of a single \( FA \), while Section~\ref{TWO_FA} examines the hydrodynamic interactions between two \( FA \)s. Finally, Section~\ref{Summary} summarizes our key findings and discusses their implications for flagellar collective motility.
\section{Experimental motivations}
\label{EXP}
In our previous work~\cite{pozveh2021resistive}, we combined high‐speed imaging, quantitative image processing, and mode analysis to characterize the wave dynamics of \(FA\)s isolated from a wall‐less \textit{C. reinhardtii} strain. Flagellar isolation was performed following the protocols of Hyams and Borisy~\cite{hyams1975flagellar,hyams1978isolated}. Figure~\ref{fig:FA_EXP}A–B shows an isolated \(FA\) reactivated with 2\,mM ATP (see Ref.~\cite{pozveh2021resistive} for experimental details). Upon reactivation, curvature waves originate at the basal ends—where the two flagella attach to the basal body—and propagate toward the distal tips (Fig.~\ref{fig:FA_EXP}C–D). The wave frequency depends on ATP concentration, following a Michaelis–Menten–type relationship~\cite{geyer2013characterization,chen2015atp,ahmad2021light,gholami2022waveform}. As the $FA$ swims, the mismatch between its two flagellar arm beat frequencies drives the basal body along an epitrochoid-like trajectory (Fig.~\ref{fig:FA_EXP}E–F), with fast oscillations at \(\sim22\ \mathrm{Hz}\) superimposed on slower undulations near \(\sim0.8\ \mathrm{Hz}\).
\textcolor{black}{\section{Regularized Stokeslet analysis}
\label{Stokselet}
\noindent In this section, using our experimentally determined $FA$ shapes as input, we apply the method of images for the regularized Stokeslet developed by Ainley \emph{et al.} \cite{ainley2008method} to compute the full three-dimensional flow field (see Appendix~\ref{Stokslets}). These Stokeslet results provide a benchmark for comparison with flow fields obtained independently via our two‐dimensional MPC simulations in Section~\ref{FA}. Although the isolated \(FA\)s swim in a shallow 3D chamber and exhibit asymmetries—different beat frequencies, mean curvatures, and flagellar lengths—we compare their computed 3D flow fields qualitatively to the 2D MPC‐derived flows of an idealized \(FA\) with perfectly symmetric arms. Experimentally, \(FA\)s were confined between two microscope slides spaced 100\,\(\mu\)m apart, and the \(FA\) in Fig.~\ref{fig:FA_EXP} was imaged at an estimated height of 20 pixels $\approx$ 9\,\(\mu\)m above the bottom slide (not measured directly). Consistent with Klindt \emph{et al.}’s boundary‐element analysis of intact \textit{C.~reinhardtii} flagella~\cite{klindt2015flagellar}, we observe that isolated \(FA\)s alternate between pusher‐like and puller‐like flow signatures, averaging to an approximately neutral swimmer (Fig.~\ref{fig:FA_EXP}G–I). However, unlike Ref.~\cite{klindt2015flagellar}, we observe that the isolated  \(FA\) acts as a pusher during the power stroke and as a puller during the recovery stroke. This difference may stem from the absence of the cell body in our isolated $FA$s, underscoring the critical impact of basal body positioning relative to the two flagellar arms and of overall flagellar morphology.}
%
\section{Simulation method}
\label{MPC}
In this work, we utilize the multiparticle collision dynamics (MPC) method~\cite{malevanets1999mesoscopic} to simulate the two-dimensional swimming dynamics of an \(FA\) in a fluid~\cite{yang2010swarm,yang2008cooperation}. The MPC technique is a well-established computational tool that has been widely applied to model the hydrodynamics of active matter and polymeric systems~\cite{mussawisade2005dynamics,gompper2009multi,khan2022effect,babu2012modeling,munch2016taylor,zantop2021multi,mandal2019multiparticle,schwarzendahl2019hydrodynamic}. In addition to solving coarse-grained Navier-Stokes equations, MPC inherently incorporates thermal fluctuations, making it a powerful tool for realistic hydrodynamic simulations. In this framework, fluid dynamics is governed by the MPC method, while the motion of the \( FA \) is described using molecular dynamics, ensuring an accurate representation of both hydrodynamic interactions and flagellar movement.
\subsection{Multi particle collision dynamics for fluid}
\label{MPC:Fluid}
In the MPC framework, the fluid is represented by point particles, each labeled by an index \( i \), with mass \( m^f \), position \( \bm{r}^f_i \), and velocity \( \bm{v}^f_i \). The fluid system is initialized by distributing these particles within a simulation box of size \( L_\text{box} \times L_\text{box} \) such that each collision cell of size \( a \times a \) contains, on average, \( \rho_0 \sim 10 \) particles, where \( a \ll L_{box} \).  Let \( \delta t \) be the time step used to update the positions and velocities of the particles. During each \( \delta t \), the system undergoes two key steps:  

1. \textit{Streaming Step}: Particles move ballistically, updating their positions according to: 
\begin{equation}
	\bm{r}^f_i(t+\delta t) = \bm{r}^f_i(t) + \bm{v}^f_i(t) \delta t. \label{stream}
\end{equation}  

\begin{figure}[t!]
	\centering
	\includegraphics[width=\columnwidth]{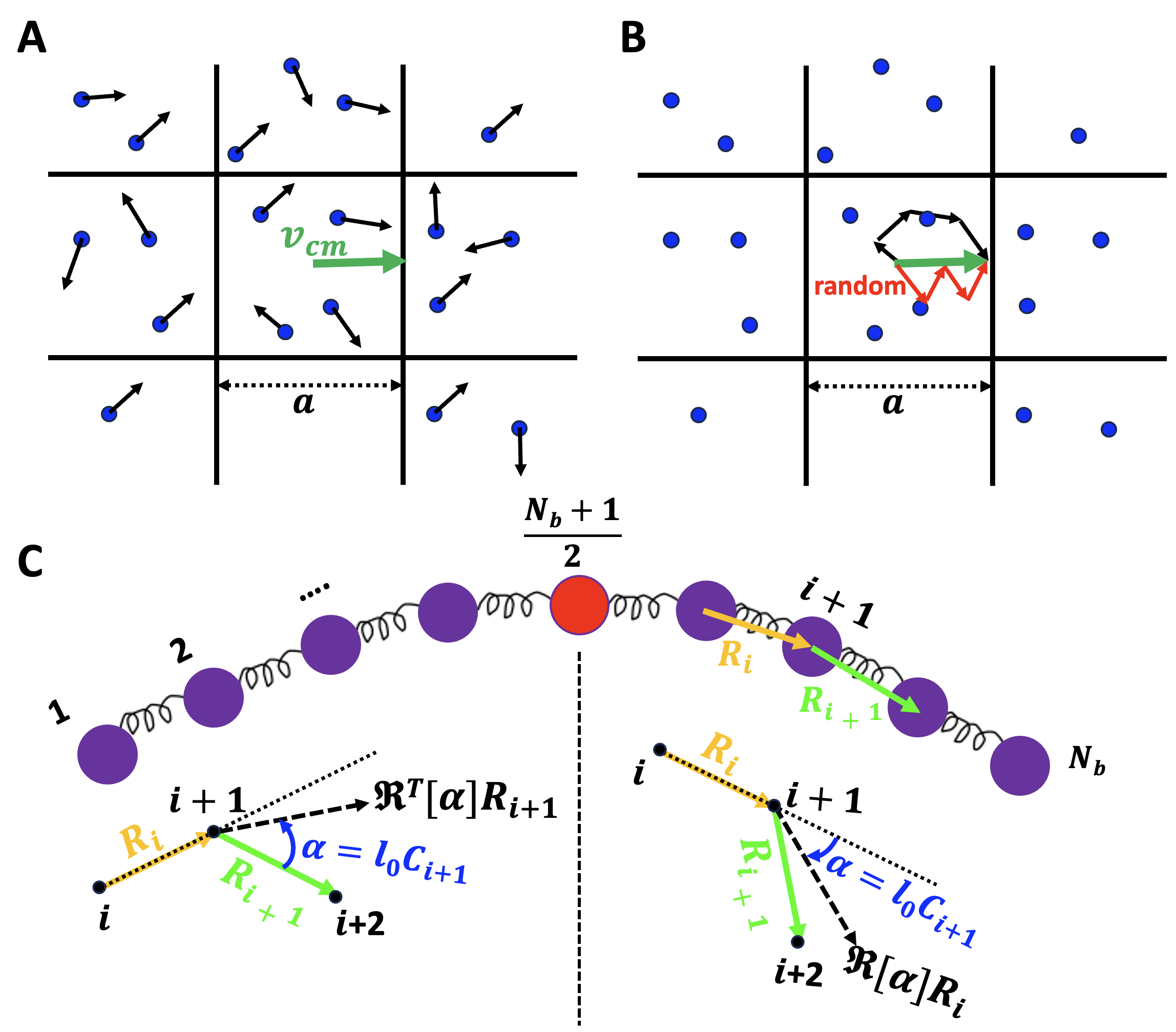}
	\caption{Schematic representation of the MPC framework.  
		(A) The simulation domain is partitioned into square cells of linear size \( a \). Each particle within a given cell possesses a velocity \( \bm{v}_i^f(t)\) (indicated by black arrows). The sum of all particle velocities in cell \( c \) determines the center-of-mass velocity \( \bm{v}_{\text{cm}}(t) \) (depicted by the dark green arrow).  
		(B) A stochastic Andersen collision step (see Eqs.~\ref{collision1}-~\ref{collision1_1}) is applied within each cell, ensuring momentum conservation. During this step, new random velocities (shown as red arrows) are assigned to the particles in cell \( c \) while maintaining the center-of-mass velocity. \color{black} (C) Schematic of the \(FA\) model: successive beads (\(N_b = 101\)) are connected by springs with spring constant \(k\) and bending potentials characterized by modulus \(\kappa\).
		}
	\label{fig:MPCD}
\end{figure}

2. \textit{Collision Step}: Particles within each collision cell interact and exchange momentum through stochastic collisions, implemented using the Andersen thermostat~\cite{gompper2009multi}. The post-collision velocities are given by:
\begin{equation}
	\bm{v}^f_i(t+\delta t) = \bm{v}^f_{cm}(t) + \bm{v}^{rand}_i(t) - \bm{v}^{rand}(t). \label{collision1}
\end{equation}  
The velocities \( \bm{v}^{rand}_i(t) \) are drawn from a normal distribution with variance \( k_B T / m^f \), ensuring a Maxwell-Boltzmann velocity distribution at equilibrium. The center-of-mass velocity of particles within a given collision cell \( c \) at time \( t \) is defined as:
\begin{equation}
	\bm{v}^f_{cm}(t) = \frac{\sum_{j \in c} m^f \bm{v}^f_j(t)}{\sum_{j \in c} m^f}.
\end{equation}
Similarly, the mean velocity of the randomly assigned post-collision velocities within the same cell is:
\begin{equation}
	\bm{v}^{rand}(t) = \frac{\sum_{j \in c} m^f \bm{v}^{rand}_j(t)}{\sum_{j \in c} m^f}.
	\label{collision1_1}
\end{equation}
The above collision rule inherently performs thermostating, maintaining a constant temperature by conserving the kinetic energy. It also ensures linear momentum conservation within each collision cell (see Fig.~\ref{fig:MPCD}A-B). However, angular momentum is not conserved by default. To restore angular momentum conservation, the post-collision velocities in Eq.~\eqref{collision1} are modified by applying a rigid-body rotation correction:
\begin{eqnarray}
	\bm{v}^f_i(t+\delta t) &=& \bm{v}^f_{cm}(t) + \bm{v}^{rand}_i(t) - \bm{v}^{rand}(t)  \nonumber \\
	& &+ m^f \Pi^{-1} \sum_{j\in c} [\bm{r}_j^f\times (\bm{v}_j^{rand} - \bm{v}_j^f)] \times \bm{r}_i^f, \label{collision2}
\end{eqnarray}
where \( \Pi \) is the moment of inertia tensor of the particles in the collision cell.

Additionally, the algorithm must be corrected for Galilean invariance, which is achieved by applying a random grid shift to the entire simulation box before each collision step~\cite{ihle2001stochastic,ihle2003stochastic}. Specifically, all particles are displaced by a random vector whose components are uniformly distributed in the interval \( [-a/2, a/2] \). After the collision step, the particles are shifted back by the same magnitude in the opposite direction. This grid shift correction is particularly necessary when the mean free path of the fluid particles is smaller than the collision cell size \( a \).
\subsection{Model for Flagellar Apparatus}
\label{MPC:FA}
\begin{figure*}[htbp!]
	\centering
	\includegraphics[width=1.0\columnwidth]{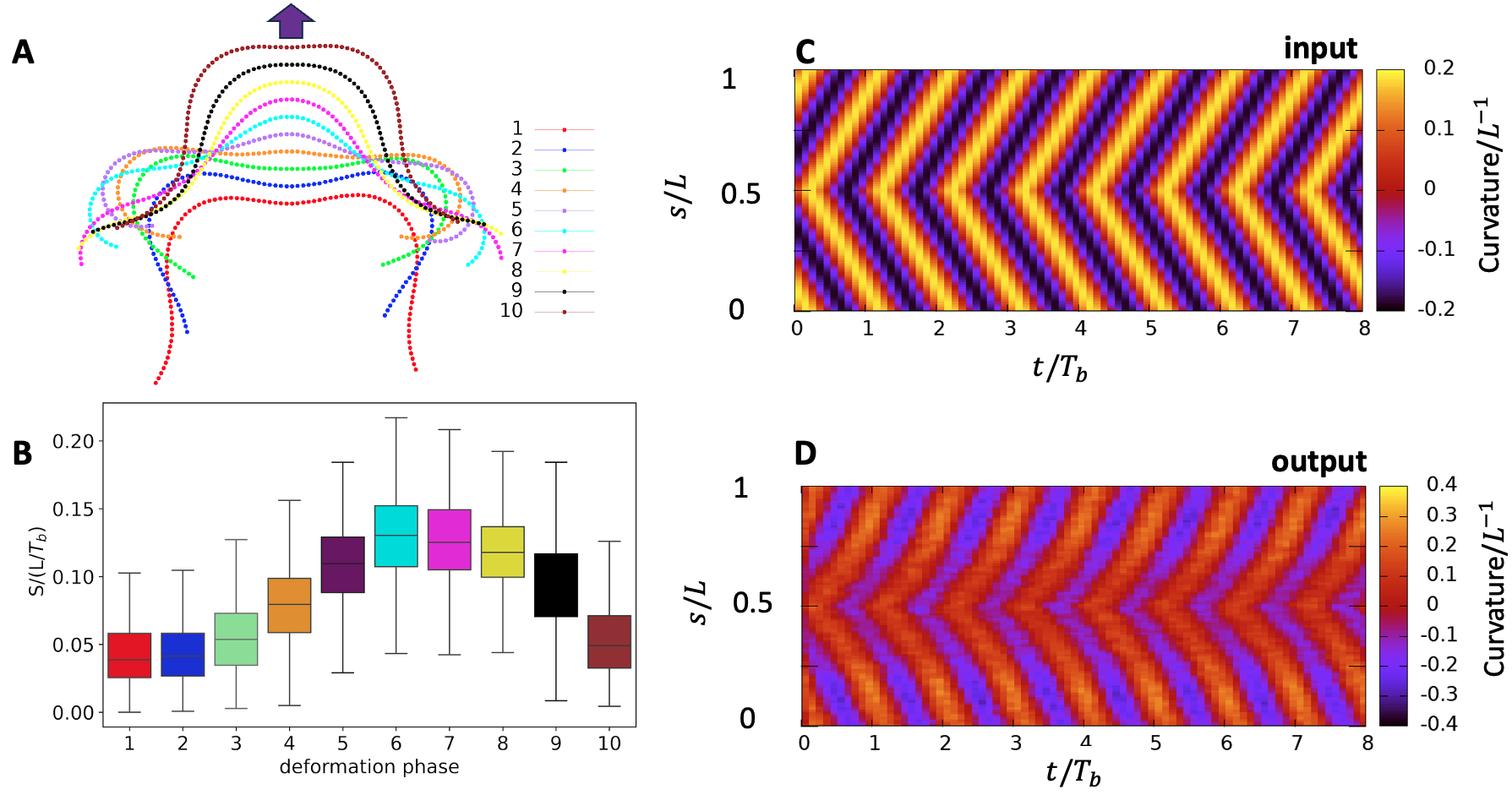}
	\caption{(A) Equidistant phases of the \( FA \) within a beat cycle. A complete sine-like wave propagates through each arm as the \( FA \) transitions from phase 1 to phase 10. Successive phases are displaced in the direction of motion of the \( FA \)  for illustration (see Video 2). The arrow indicates the direction of motion for visual reference.  (B) Speeds corresponding to the beat phases of the \( FA \). Phases \( 4 \) to \( 8 \) constitute the power stroke, while the remaining phases belong to the recovery stroke. (C) Input curvature wave along the \( FA \) as defined in Eq.~\eqref{curvewave}.  (D) Resultant curvature wave along the \( FA \), obtained from simulation data. \textcolor{black}{The resultant curvature wave closely follows the input curvature wave, apart from randomly occurring spikes caused by fluid–$FA$ interactions.}}
	\label{fig:BeatPhasesFA}
\end{figure*}
The \(FA\) is modeled as a chain of \(N_b\) monomers, each of mass \(m^b\), connected by elastic springs and bending potentials (Fig.~\ref{fig:MPCD}C), extending the framework of Ref.~\cite{yang2010swarm}.  Its total elastic energy is:
\begin{equation}
	\begin{aligned}
		E &= \sum_{i=1}^{N_b-1}\frac{k}{2l_0^2}\bigl(\|\vec R_i\|-l_0\bigr)^2 \\[-3pt]
		&\quad + \sum_{i=1}^{(N_b-1)/2}\frac{\kappa}{2l_0^3}\Bigl\|\vec R_i - \mathcal{R}^T\bigl[l_0\,C(s_{i+1},t)\bigr]\;\vec R_{i+1}\Bigr\|^2 \\[-3pt]
		&\quad + \sum_{i=(N_b-1)/2}^{N_b-2}\frac{\kappa}{2l_0^3}\Bigl\|\vec R_{i+1} - \mathcal{R}\bigl[l_0\,C(s_{i+1},t)\bigr]\;\vec R_{i}\Bigr\|^2 + V.
	\end{aligned}
	\label{EnergyOHS}
\end{equation}
The first term is the harmonic spring energy (spring constant \(k\), equilibrium length \(l_0\)), where \(\vec R_i\) is the bond vector from bead \(i\) to \(i+1\) (Fig.~\ref{fig:MPCD}C).  The second and third terms are the bending energies of the two flagellar arms (rigidity \(\kappa\)).  Here \(\mathcal{R}[\alpha]\) denotes a clockwise rotation by angle \(\alpha\) (and \(\mathcal{R}^T\) its transpose), and the local curvature:
\begin{align}
	C(s,t) = C_0
	&+ A \cos\!\Bigl[2\pi\bigl(-\tfrac{s-s_0}{\lambda}-f_1\,t+\phi_1\bigr)\Bigr]\,\Theta(s-s_0) \nonumber\\
	&+ A \cos\!\Bigl[2\pi\bigl(\tfrac{s-s_0}{\lambda}-f_2\,t+\phi_2\bigr)\Bigr]\,\bigl(1-\Theta(s-s_0)\bigr),
	\label{curvewave}
\end{align}
with \(s_0 = (N_b+1)\,l_0/2\).  In this expression, \(\Theta\) is the Heaviside step function, \(C_0\) the mean curvature, \(A\) the wave amplitude, \(\lambda\) the wavelength (taken as half the contour length), \(f_{1,2}\) the beat frequencies, and \(\phi_{1,2}\) the phase offsets.  Each bending term thus represents a sinusoidal curvature wave traveling in opposite directions along one flagellum, making the \(FA\) effectively a two‐armed microswimmer (Fig.~\ref{fig:BeatPhasesFA}).  Unless otherwise specified, we set \(\phi_1=\phi_2=0\) and \(f_1=f_2=f\) so that both arms share the same phase and frequency.

The final term in Eq.~\eqref{EnergyOHS} represents the volume‐exclusion energy \(V\), included only when simulating two interacting \(FA\)s. Steric repulsion between monomers on different \(FA\)s is modeled by a shifted, truncated Lennard–Jones potential. The total volume‐exclusion energy is:
\begin{equation}
	V \;=\; \sum_{i=1}^{N_b^{(1)}} \sum_{j=1}^{N_b^{(2)}} U(r_{ij}),
	\label{VolExPot}
\end{equation}
where the superscripts 1 and 2 denote beads on the first and second \(FA\), respectively, and:
\[
r_{ij} = \bigl\|\mathbf{r}_i - \mathbf{r}_j\bigr\|
\]
is the distance between bead \(i\) of the first \(FA\) and bead \(j\) of the second \(FA\). The pairwise potential \(U(r_{ij})\) is:
\[
U(r_{ij}) =
\begin{cases}
	4\epsilon\Bigl[\bigl(\tfrac{\sigma}{r_{ij}}\bigr)^{12} - \bigl(\tfrac{\sigma}{r_{ij}}\bigr)^{6}\Bigr] - U(r_c),
	& r_{ij} \le r_c,\\
	0,
	& r_{ij} > r_c,
\end{cases}
\]
with the potential at the cutoff distance \(r_c\) defined as:
\[
U(r_c) 
= 4\epsilon\Bigl[\bigl(\tfrac{\sigma}{r_c}\bigr)^{12} - \bigl(\tfrac{\sigma}{r_c}\bigr)^{6}\Bigr].
\]
Here, \(\epsilon = 660\,k_BT\) sets the well depth, \(\sigma = a\) is the bead diameter, and \(r_c = 2^{1/6}a\) is the cutoff radius beyond which the interaction vanishes.  

Curvature waves induce time‐dependent bending forces, rendering the \(FA\) active. During each MPC streaming step of duration \(\delta t\), the beads are advanced using a velocity‐Verlet integrator with a smaller time step \(\delta t_b = \delta t/100\), so that each fluid‐particle step comprises 100 bead sub‐steps. In the collision step, beads are sorted into MPC cells alongside fluid particles and participate in the velocity update as in Eq.~\eqref{collision2}.

All simulations were performed in MPC units: length measured in the cell size \(a\), energy in \(k_{B}T\), and mass in units of the fluid‐particle mass \(m^{f}\), giving a time unit $\tau_{\mathrm{MPC}} = a\sqrt{m^{f}/k_{B}T}.$
Periodic boundary conditions were applied in a square box of side \(L_{\mathrm{box}} = 200\,a\). The \(FA\) comprised \(N_{b} = 101\) beads of mass \(m^{b} = 10\,m^{f}\). We used an MPC streaming step of \(\delta t = 0.025\) and integrated bead motion via a velocity–Verlet scheme with a sub‐step of \(\delta t_{b} = \delta t/100\). Springs had equilibrium length \(l_{0} = a/2\) and stiffness \(k = 1.25\times10^{4}\,k_{B}T\), while bending rigidity was \(\kappa = 200\,k_{B}T\,(N_{b}-1)\,l_{0}\), ensuring a persistence length far exceeding the contour length \(L = (N_{b}-1)\,l_{0}\). Curvature waves oscillated with period \(T_{b} = 120\,\tau_{\mathrm{MPC}}\) (frequency \(f = T_{b}^{-1}\)), amplitude \(A = 0.2\,a^{-1}\), and mean curvature \(C_{0} = 0.1\,a^{-1}\). {\color{black} Ensemble averages were computed over 112 or 224 independent runs.

The kinematic viscosity of the MPC fluid with Andersen thermostat (AT) and angular‐momentum correction (MPC‐AT+a) was derived by Gompper \emph{et al.}~\cite{gompper2009multi} as:
\begin{align}
	\nu^{\mathrm{kin}}
	&= \frac{k_{B}T\,\delta t}{m_{f}}
	\Bigl(\frac{\rho_{0}}{\rho_{0} - (d+2)/4} - \tfrac12\Bigr),\nonumber\\
	\nu^{\mathrm{col}}
	&= \frac{a^{2}}{24\,\delta t}
	\Bigl(\frac{\rho_{0}-7/5}{\rho_{0}}\Bigr),\nonumber\\
	\nu &= \nu^{\mathrm{kin}} + \nu^{\mathrm{col}}.
	\label{MPCnu}
\end{align}
Substituting our simulation parameters yields \(\nu \approx 1.5\,a^2/\tau_{\mathrm{MPC}}\). Using the characteristic $FA$ size \(2R_{c}\approx 20 a\) (where \(R_{c}=1/|C_0|\)) and its mean speed \(\langle S\rangle\approx0.07\,L/T_{b}\) (see next section), the Reynolds number:
\[
Re = \frac{\langle S\rangle\,2R_{c}}{\nu} \sim 0.4
\]
remains small, in line with previous MPC studies~\cite{khan2022effect,munch2016taylor}.}
\section{ Hydrodynamics of a single flagellar apparatus }
\label{FA}
\begin{figure*}[htbp!]
	\centering
	\includegraphics[width=1.0\columnwidth]{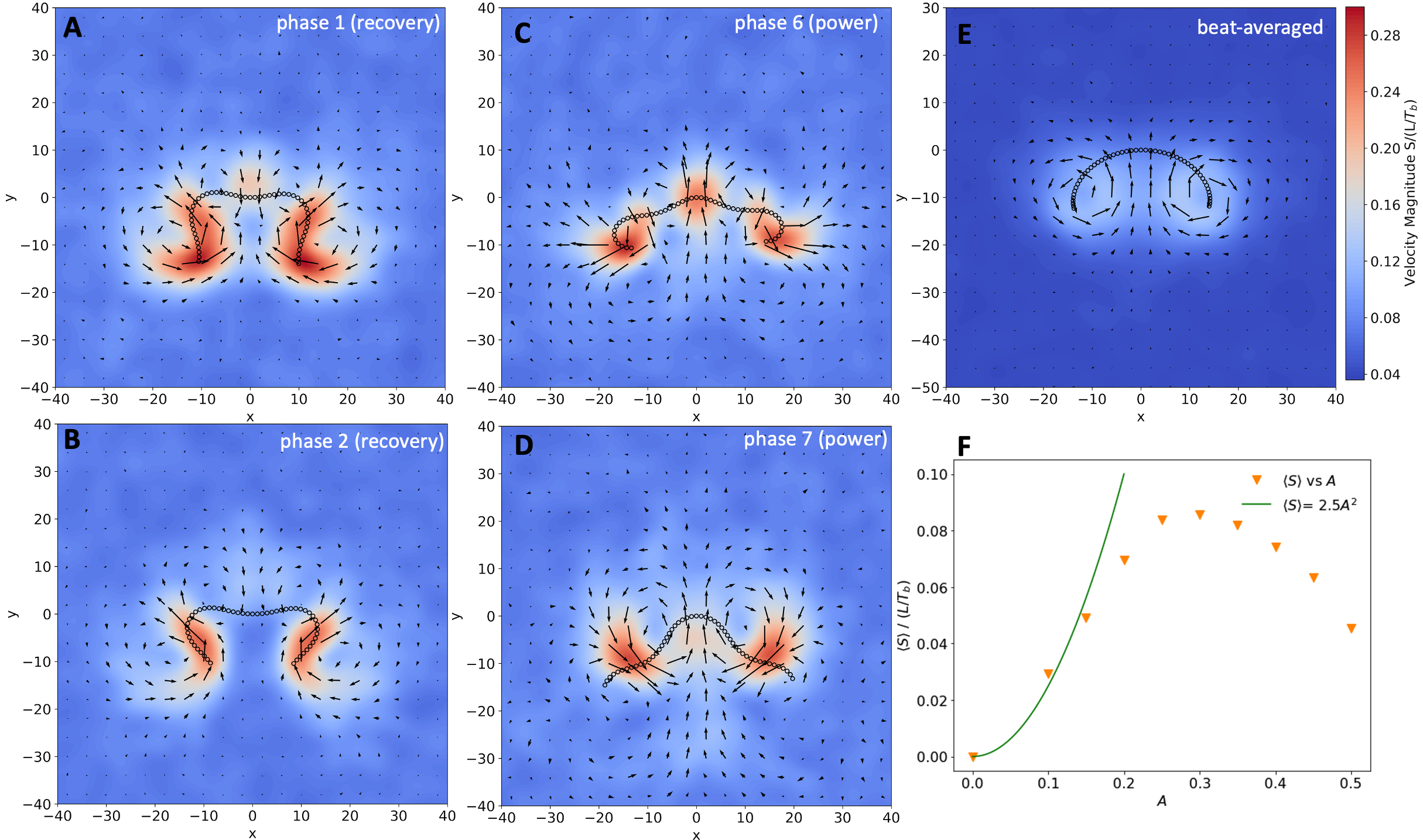}
	\caption{(A–B) Flow fields of the \(FA\) during phases 1 and 2 (recovery stroke). In both phases, flow along the axis of motion is directed opposite to the \(FA\)’s movement, and the axial magnitude is smaller than the peak values at the arm tips.  (C–D) Flow fields during phases 6 and 7 (power stroke). Here, flow along the axis aligns with the \(FA\)’s direction of motion, and its magnitude approaches the peak values. (E) Flow field of the \(FA\) averaged over one beat cycle. Along the axis of motion, the flow aligns with the \(FA\)’s direction of motion as in power stroke. The flow decays rapidly and becomes negligible beyond distances greater than twice the \(FA\)’s size (\textcolor{black}{see Video 5}).
		\textcolor{black}{(F) Time‐averaged speed \(\langle S\rangle\) versus curvature‐wave amplitude \(A\) for \(f=1/120\). At low \(A\), \(\langle S\rangle\) increases quadratically (solid fit line). Beyond a critical amplitude, speed reaches a maximum and then decreases with further increases in \(A\)}.}
	\label{fig:FArecovery}
\end{figure*}
Using the parameters and methods described in Section~\ref{MPC}, we performed MPC simulations of a flagellar apparatus under four conditions: an active \(FA\) with non-zero mean curvature \(C_0\), its inactive counterpart \(FA_i\) (by setting the curvature wave frequency \(f=0\) as shown in Video 3), an active \(FA_0\) with zero mean curvature (Video 4), and its inactive version \(FA_{i0}\). In the following, we analyze these simulations and extract key dynamical properties to compare swimming behaviors in these four cases.
\subsection{Flagellar apparatus with non-zero mean curvature ($FA$)}
\begin{figure*}[t!]
	\centering
	\includegraphics[width=1.0\columnwidth]{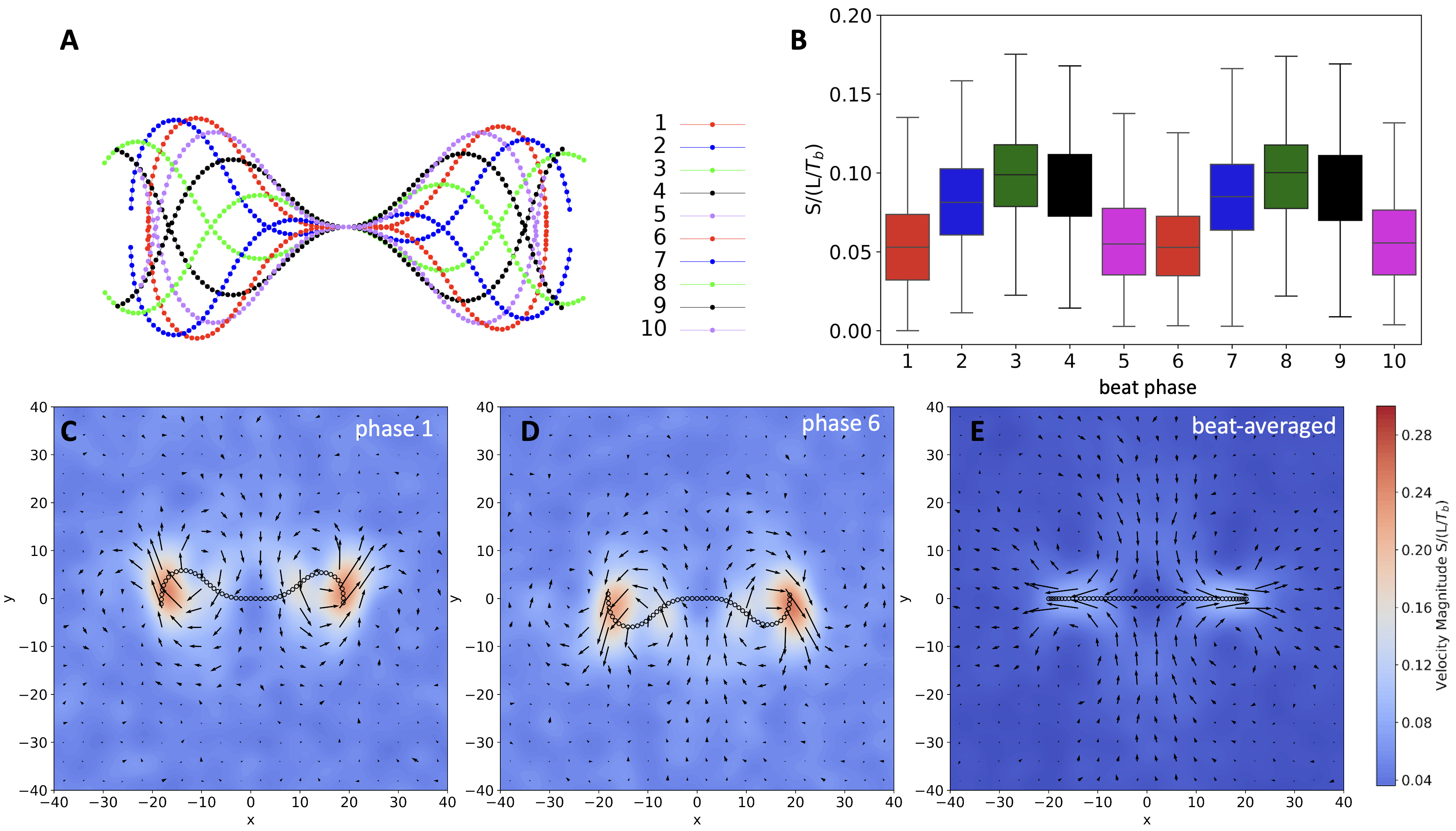}
	\caption{ (A) Equidistant phases of \( FA_0 \), with mirror images plotted in the same color.  (B) Speeds of the phases of \( FA_0 \), where the speed distributions of corresponding mirror-symmetric phases are similar; (C-D) Flow fields around the mirror-symmetric phases \( 1 \)  and \( 6 \)  of \( FA_0 \). Since the phases themselves are mirror-symmetric, their corresponding flow fields exhibit the same symmetry. Consequently, the displacement induced by phase \( 1 \) is canceled by that of phase \( 6 \).  (E) Average flow field over a beat of \( FA_0 \). The flow field is symmetric about both the \( x \)- and \( y \)-axes. In the region \( x > 0 \), the flow field resembles that of a flagellum moving to the left, while in the region \( x < 0 \), it resembles that of a flagellum moving to the right; \textcolor{black}{see Video 6}.}
	\label{fig:BeatPhasesFA0}
\end{figure*}
{\color{black} Figure~\ref{fig:BeatPhasesFA}A shows ten equally spaced snapshots of one \(FA\) beat cycle, each separated by a time interval \(T_b/10\). The beating motion is driven by the traveling curvature waves built into the Hamiltonian (Eqs.~\ref{EnergyOHS}–\ref{curvewave}). Although these input waves prescribe the \(FA\)’s curvature, the actual contour curvature arises from its interaction with the surrounding fluid. Figure~\ref{fig:BeatPhasesFA}C–D shows that the measured curvature closely matches the prescribed waveforms, with only rare, randomly occurring spikes in curvature caused by fluctuations. These two counter‐propagating, sinusoidal waves—originating at the \(FA\)’s midpoint and traveling toward each tip—generate the forward propulsion illustrated in Fig.~\ref{fig:BeatPhasesFA}A and Video 2.

Figure~\ref{fig:BeatPhasesFA}B shows the instantaneous speed distributions for each phase of the \(FA\) beat cycle, where thermal fluctuations broaden each distribution. The mean speeds during phases 4–8 (the power stroke) are slightly higher, peaking at approximately \(0.21\,L/T_{b}\) in phases 6 and 7. In contrast, phases 1–3 and 9–10 (the recovery stroke) exhibit lower mean speeds and even slight backward motion. Averaging over a full cycle gives a propulsion speed of $
\langle S\rangle \approx 0.07\,L/T_{b}$. The mean propulsion speed \(\langle S\rangle\) depends on the curvature‐wave amplitude \(A\) and frequency \(f\) defined in Eq.~\eqref{curvewave}. To explore this, we fixed \(f = 1/120\) and varied \(A\) from 0.1 to 0.5. The resulting \(\langle S\rangle\) (Fig.~\ref{fig:FArecovery}F) grows quadratically at low \(A\), reaches a maximum when undulation amplitude approaches approximately the radius of curvature of the swimmer $R_c=1/C_0\approx L/5$, and then declines—similar to the trend reported for single flagellum~\cite{khan2022effect}.

Flow fields corresponding to phases \( 1 \) and \( 2 \), which are part of the recovery stroke, are depicted in Fig.~\ref{fig:FArecovery}A-B.  Note that all flow fields presented in this study are shown in the body-fixed frame, where the frame is attached to the central bead of the $FA$, and the local tangent at this point is aligned with the \( x \)-axis. The flow fields represent the average fluid velocity in this frame, obtained after subtracting the constant background velocity corresponding to the mean fluid motion in the body-fixed frame~\cite{yang2009swimming}. The averaging is typically performed over several hundred frames. During recovery phases 1 and 2 (Fig.~\ref{fig:FArecovery}A–B), the fluid flows toward the \(FA\) along its axis—consistent with the near‐field Stokeslet-computed flow pattern in Fig.~\ref{fig:FA_EXP}G.  In contrast, during power‐stroke phases 6 and 7 (Fig.~\ref{fig:FArecovery}C–D), the flow reverses and moves away from the \(FA\) along its axis, again matching the near-field Stokeslet‐computed pattern in Fig.~\ref{fig:FA_EXP}G.  Moreover, in every beat phase, the flow along each flagellar arm exhibits two counter-rotating vortices (see Video 5) that travel from the $FA$’s base to its tip—just as seen for a single flagellum~\cite{yang2009swimming}—and these vortex pairs are also clearly visible in the Stokeslet‐computed flow fields shown in Fig.~\ref{fig:FA_EXP}G–I. 

Figure~\ref{fig:FArecovery}E presents the beat‐averaged flow field of the \(FA\), which we compare to the near‐field average shown in Fig.~\ref{fig:FA_EXP}G. The characteristic pair of vortices along each flagellar arm is also apparent in the Stokeslet‐computed average (last panel of Fig.~\ref{fig:FA_EXP}G). We attempted to extract a far‐field pattern from our MPC simulations by coarse‐graining over larger grid cells, but no distinct structure emerged. Consequently, we cannot directly compare our MPC flows to the far‐field Stokeslet result in Fig.~\ref{fig:FA_EXP}I. Although periodic boundary conditions allow correlations up to the box size, the observed flow decays on a much shorter length—likely because thermal fluctuations disrupt long‐range correlations~\cite{drescher2010direct}. Notably, the beat‐averaged flow decays even more rapidly than the individual‐phase flows, consistent with the Stokeslet average in Fig.~\ref{fig:FA_EXP}G-I. This accelerated decay probably reflects cancellation between power stroke and recovery stroke contributions beyond a distance comparable to the \(FA\) size.}

{\textcolor{black} {Finally, Ref.~\cite{pozveh2021resistive} reported that isolated \(FA\)s typically beat their two arms at different frequencies—mismatches of \(10\%\)–\(30\%\)—resulting in epitrochoid-like swimming trajectories. To test our model, we introduced a \(30\%\) frequency difference (all other parameters as in Section~\ref{MPC}) and observed an  epitrochoid-like swimming path (Fig.~\ref{fig:Helical} and Video 7), in agreement with experiment (Fig.~\ref{fig:FA_EXP}E). We note, however, that epitrochoid-like trajectories can also arise from asymmetries in mean curvature, phase offsets, or unequal arm lengths; additionally, non-planar beating patterns may produce helical swimming paths in 3D~\cite{cortese2021control,mojiri2021rapid,pozveh2021resistive}. 
}}
%
\subsection{Flagellar apparatus with zero mean curvature ($FA_0$)}
It is well known that at low Reynolds numbers, successful swimmers must exhibit non-reciprocal body kinematics. In his seminal paper, Purcell~\cite{purcell2014life} formulated the so-called \textit{Scallop theorem}, which states that if the sequence of shapes adopted by a swimmer undergoing time-periodic deformations remains identical after a time-reversal transformation, then the swimmer cannot achieve net displacement.  Mathematically, non-reciprocal kinematics is a necessary but not sufficient condition for propulsion~\cite{lauga2009hydrodynamics}. A simple counterexample is a system consisting of two flagella, each with zero mean curvature, that are mirror images of each other and are positioned head-to-head (\( FA_0 \)). Although their combined motion is non-reciprocal, the mirror symmetry of the system prevents any net displacement of the center of mass, effectively canceling out any propulsion.  In the following, we present our MPC simulation results for \(FA_0\), obtained using the same procedure and parameters as for \(FA\), except that \(C_0\) is set to zero.

Figure~\ref{fig:BeatPhasesFA0}A illustrates the various phases of \( FA_0 \) during a beat cycle. In particular, the first five phases are mirror images of the subsequent five phases, respectively. However, the mirror symmetry is slightly disrupted because of thermal fluctuations.  Similarly to the case of \( FA \), the resultant curvature wave of \( FA_0 \) closely resembles the input curvature wave. The speeds corresponding to each phase are plotted in Fig.~\ref{fig:BeatPhasesFA0}B. As in the case of \( FA \), the speeds of each phase of \( FA_0 \) exhibit a distribution that we attribute to thermal fluctuations. Notably, the speed distributions of each mirror‐image phase are nearly identical.

Flow fields around the mirror-symmetric phases \( 1 \) and \( 6 \) of \( FA_0 \), as obtained from our simulations, are shown in Fig.~\ref{fig:BeatPhasesFA0}C-D. The results indicate that the flow fields corresponding to these phases are also mirror symmetric, consistent with the symmetry of the phases themselves (see Video 6). The beat-averaged flow field of \( FA_0 \) is presented in Fig.~\ref{fig:BeatPhasesFA0}E. This flow field exhibits mirror symmetry along  the \( x \)- and \( y \)-axes. The beat-averaged flow field of a single flagellum~\cite{yang2010swarm} consists of two vortices near its rear end and an inflow perpendicular to its axis close to its head. A similar pattern can be observed in the beat-averaged flow field along each arm of \( FA_0 \) in Fig.~\ref{fig:BeatPhasesFA0}E.  Thus, the flow field of $FA_0$ can be viewed as the superposition of the fields from two head‐on connected flagella, each with zero mean curvature.

Ideally, the mirror symmetry observed in the computed flow fields and speed distributions implies that the displacement generated during the first five phases should be exactly canceled by the displacement from the subsequent five phases. However, thermal fluctuations introduce variations in the speed of each phase, resulting in a distribution of possible velocities. Consequently, perfect cancellation may not occur, leading to small random displacements.  In the next section, we compute the mean square displacement (MSD) and other related quantities to investigate the dynamics of \( FA_0 \) compared to \( FA \).
%
%
\subsection{Mean square displacement}
\begin{figure*}[t!]
	\centering
	\includegraphics[width=1.0\columnwidth]{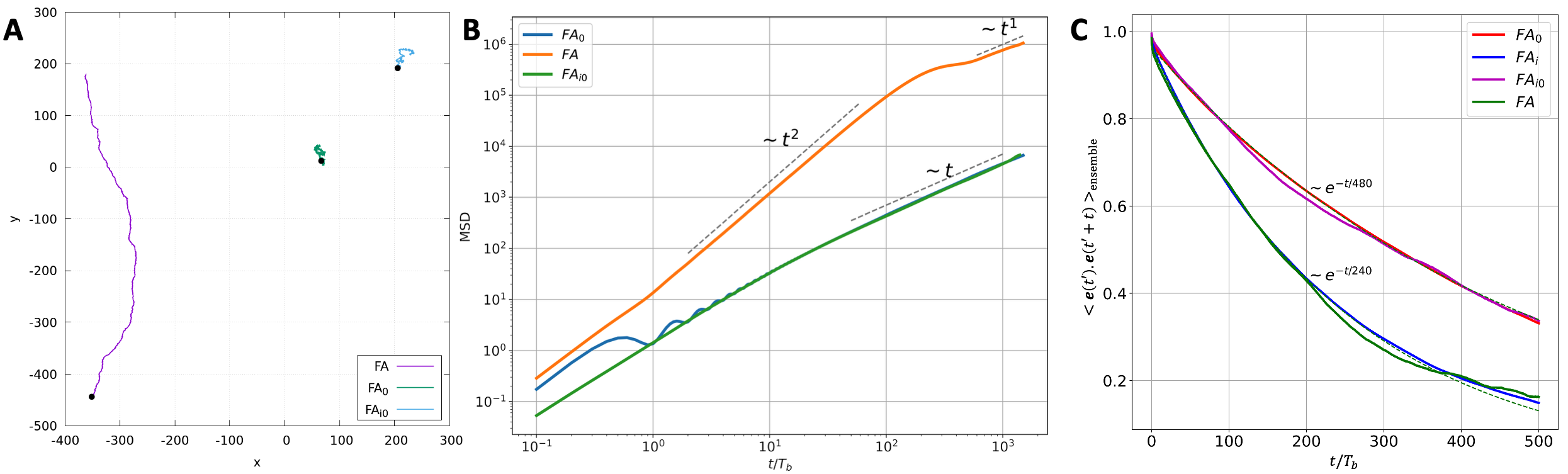}
	\caption{A) Trajectories of \( FA \), \( FA_0 \), and \( FA_{i0} \) from MPC simulations over a period of \( 200T_b \). Both \( FA_0 \) and \( FA_{i0} \) exhibit diffusive motion, while \( FA \) moves ballistically until its orientation changes due to rotational diffusion; see Videos 2-4. B) Mean square displacements of \( FA \), \( FA_0 \), and \( FA_{i0} \) obtained from the hybrid MPC simulations. Here, \( T_b \) represents the beating period. The \( FA \) exhibits ballistic motion up to approximately 240 beat periods, after which it transitions into a diffusive regime due to rotational diffusion.  In contrast, \( FA_0 \), after an initial transient phase, smoothly transitions into diffusion. The MSD curves of \( FA_0 \) and \( FA_{i0} \) overlap, indicating that \( FA_0 \) does not exhibit active swimming, thereby confirming the validity of the scallop theorem in the presence of fluctuations. C) Orientation correlation functions of \( FA \), \( FA_0 \), \( FA_i \), and \( FA_{i0} \). Dotted lines represent fitted exponential curves of the form \( e^{-t/\tau_R} \). Slight deviations at the tails of the curves arise due to finite statistical sampling. The overlap of the OACFs of \( FA \) and \( FA_i \), as well as those of \( FA_0 \) and \( FA_{i0} \), suggests that the beating activity has a negligible effect on rotational diffusion. }
	\label{fig:MSDPlot1}
\end{figure*}
The trajectories of \( FA \), \( FA_0 \), and \( FA_{i0} \) (inactive \( FA_0 \)) over a time span of 200 beats of \( FA \) are shown in Fig.~\ref{fig:MSDPlot1}A. From the figure, it is evident that \( FA \) exhibits directed swimming, covering a significantly greater distance compared to \( FA_0 \) and \( FA_{i0} \).  The \( FA \) predominantly moves in the direction indicated by the arrow in Fig.~\ref{fig:BeatPhasesFA}A and follows an approximately straight trajectory before undergoing directional changes due to reorientation, as we demonstrate below. In contrast, the distances covered by \( FA_0 \) and \( FA_{i0} \) are of the same order of magnitude, with their motion primarily governed by diffusion, as corroborated by the MSD analysis discussed below.

We calculated the MSDs using the time window method and performed an ensemble average over it, as given by: 
\begin{equation}
	\langle (\Delta \bm{x})^2 \rangle  = \frac{1}{N_e} \frac{1}{T-t} \sum_{i=1}^{N_e} \sum_{t'=0}^{T-t} (\bm x^i(t'+t) - \bm x^i(t'))^2. \label{msdSim}
\end{equation}
In the above equation, the index \( i \) represents an ensemble copy, and \( \bm x^i(t') \) denotes the position of the center of mass of the $FA$ at time \( t' \) in ensemble copy \( i \). Here, \( N_e \) is the total number of ensemble copies which is $224$ for current study and \( T \) is the total simulation time. 

The MSDs of \( FA \) and \( FA_0 \) are plotted as a function of time in Fig.~\ref{fig:MSDPlot1}B. For comparison, the MSD of \( FA_{i0} \) is also included.  The MSD curve of \( FA \) exhibits a growth of approximately \( \sim t^2 \), indicating ballistic motion up to around 240 beat periods. Beyond this point, the MSD transitions into a diffusive regime with \( MSD \sim t \). This crossover occurs as the orientation of \( FA \) changes due to rotational diffusion.

 The MSD curve of \( FA_{i0} \) follows a \( \sim t \) behavior after an initial transient of approximately ten beats, as expected for thermal diffusion. Similarly, the MSD of \( FA_0 \) also tends to \( \sim t \) after the initial transient, indicating that the system is unable to swim and instead undergoes pure diffusion driven by thermal fluctuations. Furthermore, we do not observe any enhancement in the diffusion coefficient of \( FA_0 \) compared to that of \( FA_{i0} \) due to its beating activity. This finding aligns with the results in Ref.~\cite{lauga2011enhanced}, which states that diffusion enhancement occurs only in the regime where \( \omega \tau_R \sim 1 \), whereas the current study corresponds to \( \omega \tau_R \gg 1 \). Here, \( \omega = 2\pi/T_b \) represents the angular frequency of flagellar beating, and \( \tau_R \) is the relaxation time for rotational diffusion, which we discuss in the next section.  
\subsection{Orientation auto-correlation function}
 We observed that the \(FA\)’s mean‐squared displacement (MSD; Fig.~\ref{fig:MSDPlot1}B) transitions from a ballistic regime to a diffusive one—an effect we attribute to rotational diffusion. To quantify this, we compute the orientation autocorrelation function (OACF). At each time \(t\), we extract the tangent vector \(\bm{e}(t)\) at the $FA$’s midpoint and evaluate
 \[
 \mathrm{OACF}(t) \;=\; \bigl\langle \bm{e}(t')\!\cdot\!\bm{e}(t'+ t)\bigr\rangle,
 \]
where \(\langle \cdot \rangle\) denotes an ensemble average. The OACFs of \( FA \) and \( FA_0 \) are shown in Fig.~\ref{fig:MSDPlot1}C. For comparison, we also computed the OACFs of \( FA_i \) and \( FA_{i0} \) and included them in Fig.~\ref{fig:MSDPlot1}C.   Interestingly, the OACFs of \( FA \) and \( FA_i \) overlap, and similarly the OACFs of \( FA_0 \) and \( FA_{i0} \) match. This suggests that the beating activity does not significantly affect the rotational diffusion constant of the $FA$s.
 
 The OACFs of both \(FA\)s fit well to exponential functions of the form \(e^{-t/\tau_{R}}\), as shown by the dotted lines in Fig.~\ref{fig:MSDPlot1}C. We find \(\tau_{R}\approx240\,T_{b}\) for \(FA\) and \(\tau_{R}\approx480\,T_{b}\) for \(FA_{0}\), with \(T_{b}\) the beat period. The value of \(\tau_{R}\) for \(FA\) matches the MSD crossover to diffusion, indicating that rotational diffusion drives this transition. In particular, these \(\tau_{R}\) values are an order of magnitude smaller than the theoretical predictions, as discussed below.
\begin{figure}[t!]
	\centering
	\includegraphics[width=1.0\columnwidth]{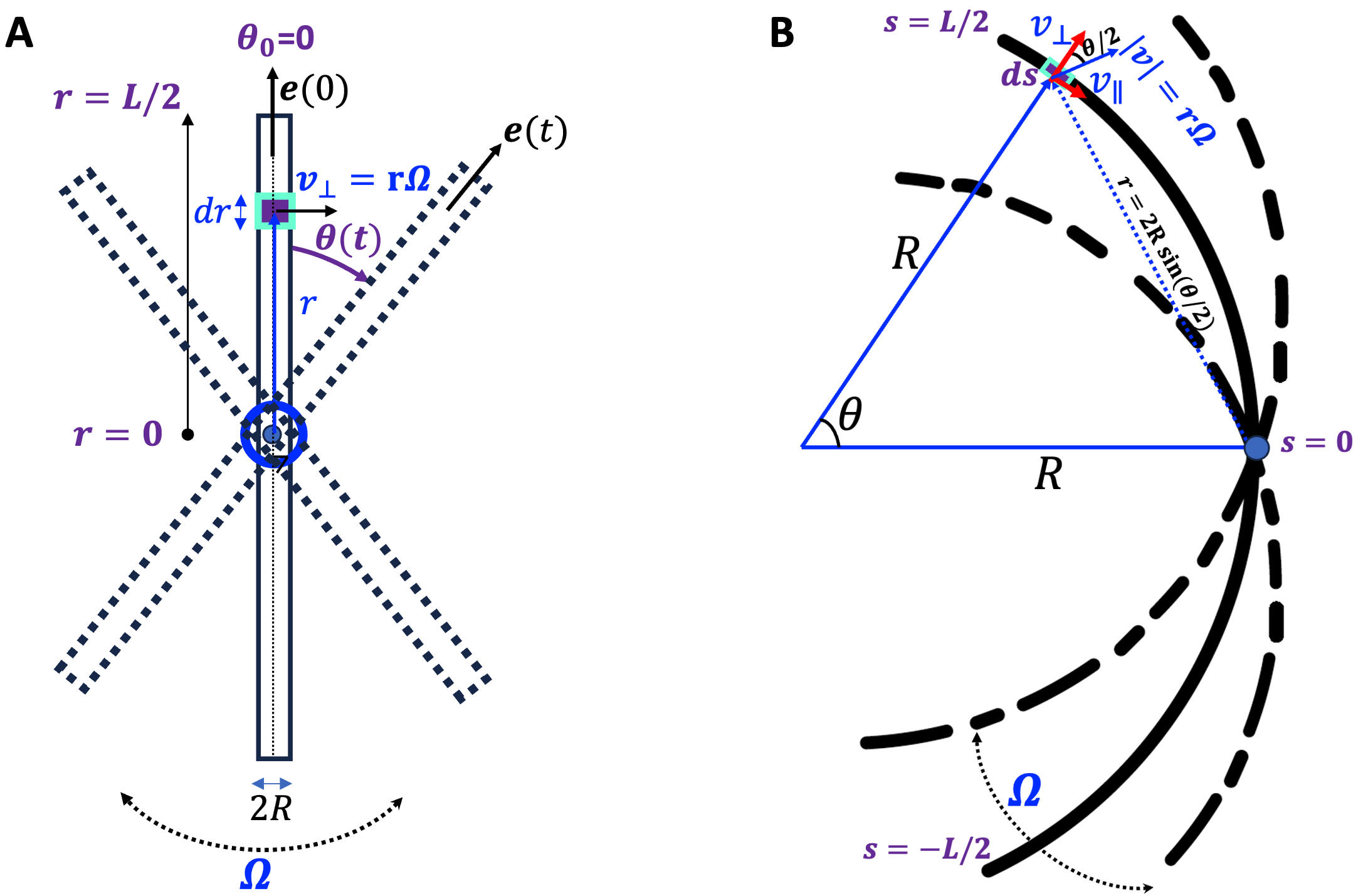}
	\caption{(A) A schematic representation of a slender, rigid cylinder of length \( L \) rotating in the \( x \)-\( y \) plane about a point passing through its center.  
		(B) A bent, slender cylinder with mean curvature \( C_0 = 1/R \) rotating about its center, marked by a small blue circle. }
	\label{fig:Rod}
\end{figure}

Theoretically, we can estimate \( \tau_R \) for a slender, rigid cylinder immersed in a fluid of viscosity \( \mu \).  Consider a slender cylinder of length \( L \) and radius \( R \) lying in the \( x \)-\( y \) plane. For convenience, we assume that at \( t=0 \), the cylinder is oriented at an angle \( \theta_0 = 0 \), measured from the \( +y \)-axis. The characteristic time scale \( \tau_R \) for this rigid rod to lose its orientation is related to the rotational diffusion constant \( D_R \) as:
\begin{equation}
	\sigma_\theta^2 = \left\langle (\theta(t) - \theta_0)^2 \right\rangle = \left\langle \Delta \theta(t)^2 \right\rangle = 2D_R t.
\end{equation}
Next, we examine the correlation of the direction vector:
\begin{equation}
	\left\langle \bm{e}(t) \cdot \bm{e}(0) \right\rangle = \left\langle \cos(\theta - \theta_0) \right\rangle = \left\langle \cos(\theta) \right\rangle = \int_{-\infty}^{+\infty} e^{i\theta} p(\theta) d\theta,
\end{equation}
where \( p(\theta) \) follows a normal distribution:
\begin{equation}
	p(\theta) = \frac{1}{\sqrt{2\pi\sigma_\theta^2}} e^{-\frac{\theta^2}{2\sigma_\theta^2}}.
\end{equation}
Evaluating the integral, we obtain the following:
\begin{equation}
	\left\langle \cos\theta \right\rangle = e^{-D_R t}.
\end{equation}
%
This result indicates that \( \tau_R= 1/D_R\) represents the time it takes for the correlation to drop by a factor of \( 1/e \).  
To relate \( D_R \) to the transverse drag coefficient \( \eta_\perp \) of the cylinder, we consider a scenario where the cylinder rotates at instantaneous angular velocity \( \Omega \) about an axis perpendicular to the \( x \)-\( y \) plane (along the \( z \)-direction), passing through its center (see Fig.~\ref{fig:Rod}A). The drag torque \( \Gamma_{\text{D}} \) acting on the cylinder is proportional to the angular velocity \( \Omega \), with the rotational drag coefficient \( \kappa_{D} \). We express \( \kappa_{D} \) in terms of the transverse drag coefficient \( \eta_\perp \) by calculating \( \Gamma_{\text{D}} \) as:
\begin{align}
\Gamma_{\text{D}}^\text{straight}&=2\int_0^{L/2} d\Gamma_{\text{D}}=2\int_0^{L/2}  r(f_\perp dr)=-2\int_0^{L/2} r(\eta_\perp v_\perp)dr\nonumber\\ &=-2\int_0^{L/2} r(\eta_\perp r\Omega)dr=-2\eta_\perp \Omega\int_0^{L/2} r^2 dr\nonumber\\ &=-\frac{1}{12}L^3\eta_\perp\Omega\approx-0.08L^3\eta_\perp\Omega,
\end{align}
which gives $\kappa_{D}=L^3\eta_\perp/12$. Note that in the equation above, \( f_\perp \) represents the drag force per unit length, \( v_\perp = r\Omega \) is the instantaneous normal velocity of a small radial segment \( dr \), and the transverse drag coefficient is given by \( \eta_\perp = \frac{4\pi\mu}{\ln(L/R) - 0.5} \)~\cite{gray1955propulsion}. Using the Einstein-Smoluchowski relation, we obtain:
\begin{equation}
	D_R=\tau_R^{-1}=\frac{k_B T}{\kappa_{\text{D}}}=\frac{12 k_B T}{L^3\eta_\perp}.
\end{equation}
Here, the dynamic viscosity \(\mu\) is given by \(\mu = \nu \rho\), where the kinematic viscosity \(\nu\) is the sum of two contributions, the kinetic viscosity \(\nu_{\text{kin}}\) and the collision viscosity \(\nu_{\text{col}}\). It is measured in units of \( a_0^2 / \tau_{\text{MPC}} \) and, with our chosen parameters, equals \( 1.5 \)~\cite{gompper2009multi}.   \(\rho\) is the average particle number in each box, given by \( 10 \, m_f / a^2 \).

Next, we repeat the calculation for a slender cylinder of length \( L \) with mean curvature \( C_0 = 1/R \), where \( R \) is the radius of curvature (see Fig.~\ref{fig:Rod}B)
	\begin{align}
		\Gamma_D^\text{bent} 
		&= 2 \int_0^{L/2} \mathbf{r} \times \mathbf{f} \, ds 
		= 2 \int_0^{L/2} r \frac{\mathbf{f} \cdot \mathbf{v}}{\lvert \mathbf{v} \rvert} \, ds
		= -2 \int_0^{L/2} r \, \frac{\eta_\parallel v_\parallel^2 + \eta_\perp v_\perp^2}{\lvert \mathbf{v} \rvert} \, ds 
		\nonumber\\
		&= -2 R \, \eta_\perp \int_0^{\tfrac{L}{2R}} r \, d\theta 
		\frac{0.5 \, r^2 \Omega^2 \sin^2\!\bigl(\tfrac{\theta}{2}\bigr) 
			+ r^2 \Omega^2 \cos^2\!\bigl(\tfrac{\theta}{2}\bigr)}{r \, \Omega} 
		\nonumber\\
		&= -4 R^3 \Omega \, \eta_\perp 
		\int_0^{\tfrac{L}{2R}} d\theta \,
		\sin^2\!\bigl(\tfrac{\theta}{2}\bigr) 
		\Bigl(1 + \cos^2\!\bigl(\tfrac{\theta}{2}\bigr)\Bigr) 
		\nonumber\\
		&= -\frac{4}{125} \, L^3 \Omega \, \eta_\perp 
		\int_0^{\tfrac{5}{2}} d\theta \,
		\sin^2\!\bigl(\tfrac{\theta}{2}\bigr) 
		\Bigl(1 + \cos^2\!\bigl(\tfrac{\theta}{2}\bigr)\Bigr)  
		\nonumber\\
		&\approx -0.04 \, L^3 \, \eta_\perp \, \Omega.
	\end{align}
	Here, 	\(\mathbf{f} = -\eta_\perp \, v_\perp \, \mathbf{e}_\perp - \eta_\parallel \, v_\parallel \, \mathbf{e}_\parallel\) and 
	\(\mathbf{v} = v_\perp \, \mathbf{e}_\perp + v_\parallel \, \mathbf{e}_\parallel\), where $\mathbf{f}$ is the drag force per unit length, and \( v_\perp \) and \( v_\parallel \) denote the perpendicular and tangential velocity components of the segment \( ds \), respectively.
	We take \(\eta_\parallel=\eta_\perp/2\)~\cite{gray1955propulsion,gray1964locomotion} and \( C_0 = 5 L^{-1} \), which is the value used in our simulations. For simplicity, we have ignored the correction to \(\eta_\perp\) that arises from the mean curvature of the cylinder.

Using the MPC parameters, we find that \( \tau_D \) for a bent cylinder with mean curvature \( C_0 = 5 L^{-1} \) is approximately half of the value for a straight cylinder, which is consistent with our simulation results (see Fig.~\ref{fig:MSDPlot1}C). However, the theoretical estimates are an order of magnitude larger than those obtained from simulations. We attribute this discrepancy to the fact that the finite thickness of \( FA \) is completely neglected in the current MPC model, where the two flagella are represented by connected beads of zero physical size. Experimentally, the radius of a flagellum is approximately \( 100 \) nm, which corresponds to about \( 1\% \) of the length of each flagellum. In our MPC simulations, the length of a single flagellum was taken as \( 25a \) (where \( a \) is the lattice size). To address this discrepancy, we believe that future MPC simulations should incorporate beads of finite diameter of at least \( a/2 \)~\cite{gotze2010mesoscale}.
%
%
\section{Interaction between two $FA$s}
\label{TWO_FA}
{\textcolor{black}{Microswimmers often form swarms and exhibit collective motion, enabling efficient navigation through viscous fluids~\cite{moore2002exceptional,immler2007hook,gray1964locomotion}. Hydrodynamic interactions underlie this collective behavior. To explore these interactions, we simulated two \(FA\)s with identical physical parameters but varying curvature wave phases. We conducted two sets of numerical experiments:
\begin{enumerate}
	\item \textbf{Set I:} Both \(FA\)s were aligned to swim in the same direction along the same axis, with the second \(FA\) initially following the first.
	\item \textbf{Set II:} The \(FA\)s were aligned to swim in opposite directions along the same axis.
\end{enumerate}}

{\textcolor{black}{ In Set \(\mathbf{I}\), \(FA\)s started at a separation of \(6a\) and swam for about 100 beat periods. We fixed the phase of one $FA$ to zero and varied the other so that $\Delta\phi/2\pi$ spans \([0,1]\). Our simulations show that the two \(FA\)s form parallel-aligned bound pairs that move together when \(\Delta \phi / 2\pi \in [0,0.55]\) and again when \(\Delta \phi / 2\pi \in [0.75,1]\) (see Videos 8). As seen in studies of sperm and flagella~\cite{yang2008cooperation,yang2010swarm}, the \(FA\)s attract each other and synchronize their beating in a few tens of cycles as they pair up. When \(\Delta \phi / 2\pi = 0\), the two \(FA\)s are symmetrically aligned and beat in perfect synchrony once paired. The parallel‐aligned pair then swims in the same direction at a speed slightly higher than that of a single \(FA\). Figure~\ref{fig:FADistPD}A and Video 9 show both \(FA\)s fully aligned when \(\Delta \phi / 2\pi = 0\). Panel A of Fig.~\ref{fig:FlowFieldFAnT2} displays the flow field averaged over one beat cycle for this synchronized pair. Although the flow pattern closely resembles that of a single \(FA\) (see Fig.~\ref{fig:FArecovery}E), the magnitudes are slightly larger because the two swimmers act in unison.}}
		
{\textcolor{black}{As \(\Delta \phi/2\pi\) increases from \(0\) to \(0.55\), one \(FA\) tilts so that one of its arms beats in synchrony with the corresponding arm of the other. Figures~\ref{fig:FADistPD}B and C show snapshots of the \(FA\) pair at \(\Delta \phi/2\pi = 0.25\) and \(0.5\), respectively (see Videos 8-9). In Panel B, the bottom $FA$ tilts left, bringing its right arm into phase with the top $FA$’s right arm; in Panel C, it tilts right, synchronizing its left arm with the top $FA$’s left arm. Each oblique alignment—and its mirror image, also seen in simulations—yields a curved swimming trajectory. As \(\Delta \phi/2\pi\) exceeds 0.55, the misalignment grows and the centers of mass drift apart; for \(\Delta \phi/2\pi \in [0.6,\,0.7]\), co‐moving pairs become unstable and separate. For \(\Delta \phi/2\pi \in [0.75,1.0]\), the \(FA\)s form co‐moving bound pairs again. Figure~\ref{fig:FADistPD}F shows a snapshot of the $FA$ pair for \(\Delta \phi/2\pi = 0.75\), where the pair is more symmetrically aligned than at \(\Delta \phi/2\pi = 0.25\) or \(0.5\), although the phase difference remains visible in their shapes. }}
		
{\textcolor{black}{Figure~\ref{fig:FADistPD}G plots the distance between the centers of mass, \(d_{\mathrm{COM}}\), of the \(FA\) pair as a function of \(\Delta \phi/2\pi\), summarizing how formation of pair depends on the phase difference. \(d_{\mathrm{COM}}/L \ll 1\) indicates that hydrodynamic attraction has led to the formation of bound pairs. For \(\Delta \phi/2\pi < 0.15\), \(d_{\mathrm{COM}}\) remains nearly constant at a value close to the minimum of the Lennard–Jones potential used in Eq.~\eqref{VolExPot}. In the interval \(\Delta \phi/2\pi \in [0.15,\,0.3]\), \(d_{\mathrm{COM}}\) grows roughly linearly and then saturates for \(\Delta \phi/2\pi \in (0.3,\,0.55]\). For \(\Delta \phi/2\pi \in (0.55,\,0.7]\), \(d_{\mathrm{COM}}/L\gg 1\), indicating that there is no pair formation. Finally, when \(\Delta \phi/2\pi \in [0.75,\,1.0]\), \(d_{\mathrm{COM}}/L \ll 1\) and decreases linearly as \(\Delta \phi/2\pi\) increases, reflecting a more symmetric alignment of the \(FA\) pair.}}
\begin{figure}[t!]
			\centering
 			\includegraphics[width=\linewidth]{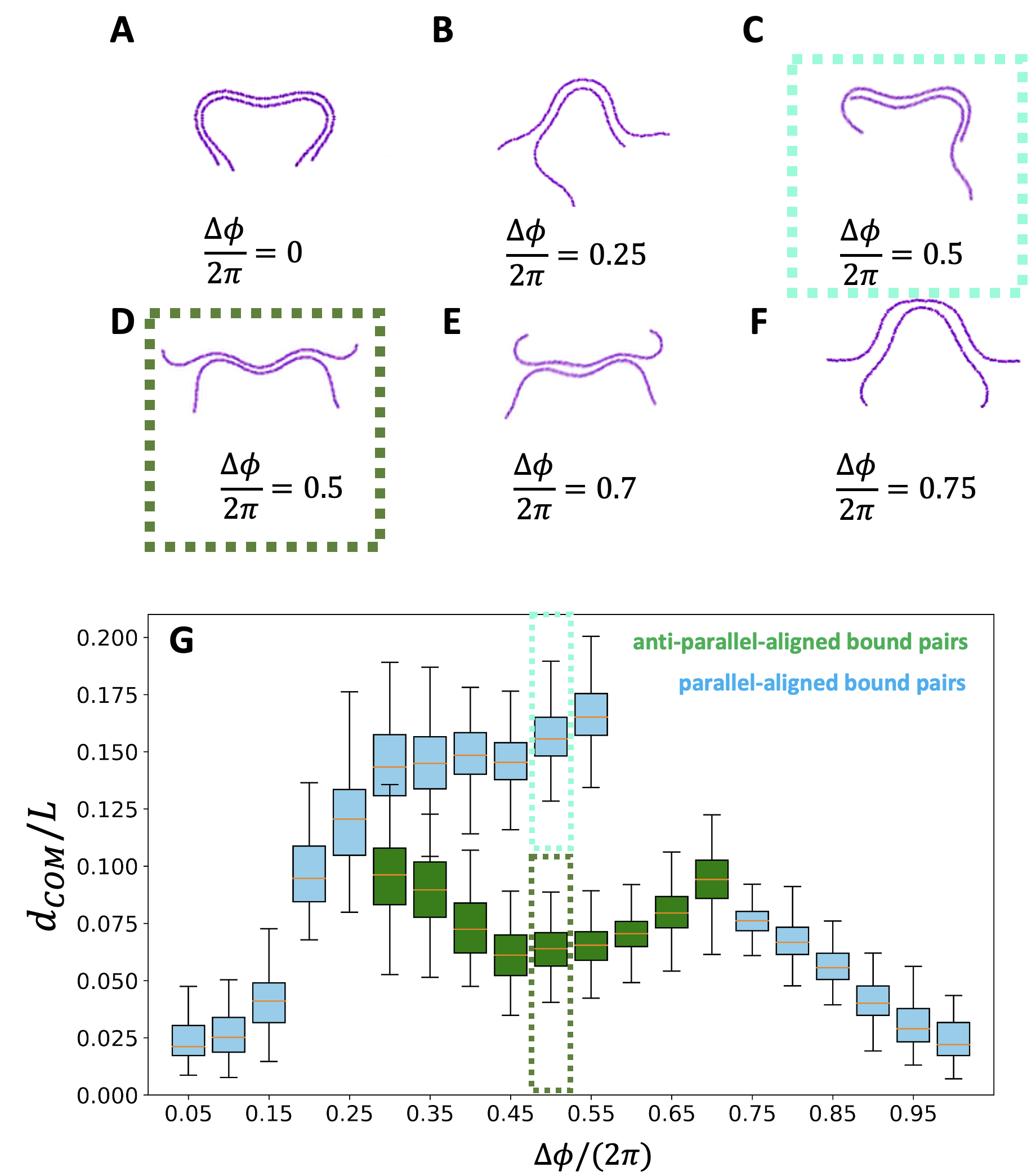}
			\caption{{\color{black} A-F) Shapes of $FA$ pairs at different phase offsets, illustrating both parallel and anti-parallel alignment configurations. (G) Distance between the centers of mass of the two bound $FA$s, plotted as a function of $\Delta\phi/2\pi$. Note that two distinct pairing configurations are possible when $\Delta\phi/2\pi\in[0.3,0.55]$; see Videos 8-11}.}
			\label{fig:FADistPD}
		\end{figure}
		\begin{figure*}[t!]
			\centering
			\includegraphics[width=1\columnwidth]{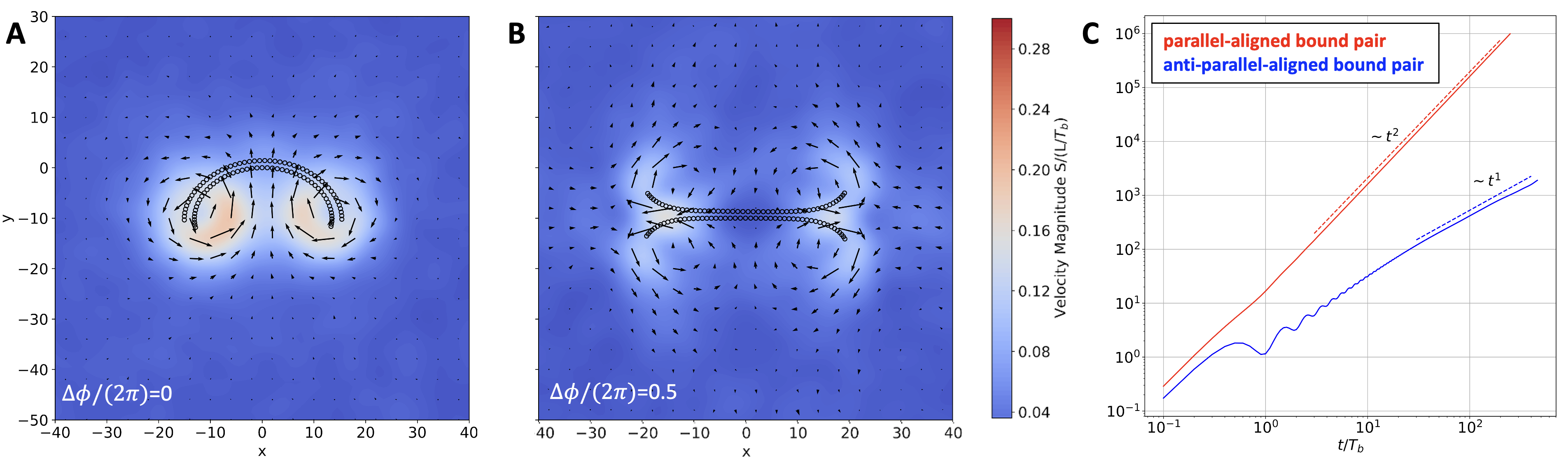}
			\caption{Beat‐averaged flow fields of bound \(FA\) pairs:  
				(A) Parallel‐aligned pair at \(\Delta\phi/2\pi = 0\) swims ballistically (see panel C).  
				(B) Anti‐parallel‐aligned pair at \(\Delta\phi/2\pi = 0.5\) mutually obstructs and exhibits diffusive motion.  
				The flow in (A) closely resembles that of a single \(FA\), but with slightly higher magnitudes (see Videos 9,11).}
			\label{fig:FlowFieldFAnT2}
		\end{figure*}
		\begin{figure*}[htbp!]
			\centering
			\includegraphics[width=1\columnwidth]{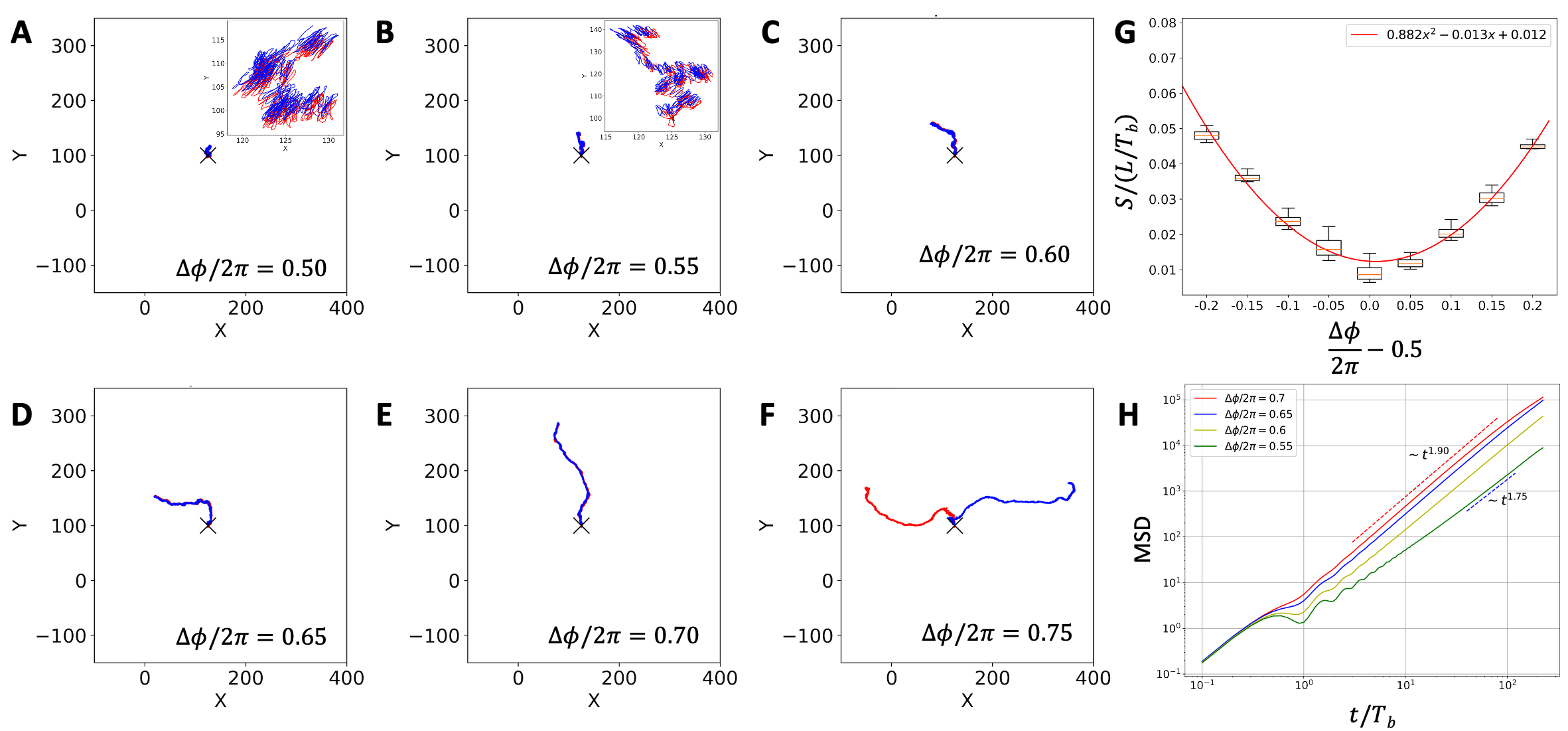}
			\caption{{\color{black} Trajectories, speeds, and MSDs of anti‐parallel‐aligned bound \(FA\) pairs. Panel A shows a Brownian‐like trajectory at \(\Delta\phi/2\pi = 0.5\). Panels B–E display progressively directed swimming as \(\Delta\phi/2\pi\) moves away from 0.5, while Panel F illustrates pair instability and separation at \(\Delta\phi/2\pi = 0.75\). The initial position of the $FA$ pair is indicated by a cross. The box plot in Panel G, plotted against \(\lvert\Delta\phi/2\pi - 0.5\rvert < 0.25\), reveals a quadratic increase in speed with phase offset. MSD curves in Panel H exhibit  power‐law exponents near 2, confirming (near‐)ballistic motion for these bound pairs.	}}
			\label{fig:nT2TrajOppPD}
		\end{figure*}

		 {\textcolor{black}{In Set \(\mathbf{II}\), with the two \(FA\)s swimming in opposite directions, we began with an initial separation of \(6a\), held one $FA$’s curvature phase fixed at \(\phi=0\), and varied the other so that $\Delta \phi/2\pi\in[0,1]$. Simulations show that the \(FA\)s form anti-parallel bounds pairs when \(\Delta \phi / 2\pi \in [0.3,\,0.7]\) (see Video 10). Figure~\ref{fig:FADistPD}D illustrates an anti-parallel "blocking" pair at \(\Delta \phi / 2\pi = 0.5\), which is perfectly symmetric about their shared axis of motion. Video 11 shows that while one $FA$ performs its power stroke, the other undergoes recovery, resulting in no net displacement—the pair simply oscillates back and forth in place. This lack of ballistic motion is reflected in the MSD, which grows linearly with time (\(t\)) as shown in Fig.~\ref{fig:FlowFieldFAnT2}C. Panel B of Fig.~\ref{fig:FlowFieldFAnT2} shows the beat‐averaged flow field of the pair. Due to symmetry, opposing flows along the axis cancel near the center, while edge flows remain directed outward in opposite directions, resulting in no net swimming. For \(\Delta \phi/2\pi\) in \([0.3,\,0.7]\) but not equal to 0.5, anti‐parallel-aligned bound pairs still form, albeit with a slight tilt. Figure~\ref{fig:FADistPD}E shows the \(\Delta \phi/2\pi=0.7\) snapshot, where this asymmetry produces a small net propulsion. In Figure~\ref{fig:FADistPD}G, \(d_{\mathrm{COM}}\) reaches its minimum at \(\Delta \phi/2\pi=0.5\) and increases symmetrically as \(\lvert\Delta \phi/2\pi-0.5\rvert\) grows. Outside this interval, no anti‐parallel-aligned bound pairs form and \(d_{\mathrm{COM}}\gg1\).
		 }}

	 {\textcolor{black}{As noted above, anti‐parallel-aligned pairs do not perfectly block each other once the phase difference deviates from \(\pi\). Figure~\ref{fig:nT2TrajOppPD} shows trajectories of anti‐parallel-aligned bound pairs over hundred beat periods for selected \(\Delta \phi/2\pi\) values. At \(\Delta \phi/2\pi = 0.5\) (panel A), the trajectory resembles Brownian motion—indicating only diffusion, with no net swimming, as confirmed by the MSD in Fig.~\ref{fig:FlowFieldFAnT2}C. For \(\Delta \phi/2\pi = 0.55\) (panel B), a slight zigzag motion is apparent, yet the pair still travels a significant distance over a hundred beats. Panels C–E display clear swimming‐like trajectories. When \(\Delta \phi/2\pi = 0.75\), the pair becomes unstable and the two \(FA\)s separate. In Fig.~\ref{fig:nT2TrajOppPD}G, the pair’s speed versus \(\Delta \phi\) increases roughly quadratically for \(\lvert \Delta \phi/2\pi - 0.5\rvert < 0.25\). Moreover, MSDs plotted in Fig.~\ref{fig:nT2TrajOppPD}H exhibit power‐law exponents near 2 for non-zero \(\lvert \Delta \phi/2\pi - 0.5\rvert\), confirming (near-)ballistic motion.}}

{\textcolor{black}{In all of the above simulations, we used an MPC fluid with average density \(\rho_{0}=10\). Theers \emph{et al.}~\cite{theers2018clustering} showed that at such low densities, the fluid segregates into low‐density regions around microswimmer clusters and high‐density zones devoid of swimmers, even forming empty “void” cells. They attributed this to spurious density segregation and recommended increasing \(\rho_{0}\) to 80—at the cost of a higher computational time. In our own tests (Appendix~\ref{clustering}), we found that this artifact appears only when the number of swimmers \(N\gg10\). For \(N\lesssim O(10)\), the fluid remains homogeneous with no artificial segregation. Because our simulations include fewer than ten swimmers, the observed attraction between two $FA$s must arise from genuine hydrodynamic interactions rather than numerical artifacts.
 }}
\section{Summary}
\label{Summary}
We developed a two‐armed microswimmer model inspired by the flagellar apparatus (\(FA\)) of \textit{C. reinhardtii} and used multiparticle collision dynamics (MPC) to analyze its swimming dynamics and hydrodynamic interactions. The \(FA\) is represented as a chain of beads linked by elastic springs with both stretching and bending stiffness. Propulsion is driven by sinusoidal curvature waves imposed along the chain, which originate at the midpoint and travel toward each tip in opposite directions, mimicking the natural flagellar undulations.

{\color{black} Flow‐field analyses of our MPC simulations show that an \(FA\) alternates between pusher‐ and puller‐type signatures—matching our 3D Stokeslet results—yet averages out as a near‐neutral swimmer. This behavior aligns with Ref.~\cite{klindt2015flagellar}, with one key difference: both MPC and Stokeslet analyses indicate that the \(FA\) acts as a puller during the recovery stroke and as a pusher during the power stroke. We attribute this difference to the absence of the cell body in our isolated $FA$, highlighting that basal body positioning, the spatial arrangement of the two flagella, and their anchoring geometry critically influence swimming dynamics.
	
Our MPC simulations show that an \(FA\) with non-zero mean curvature swims effectively: its mean‐squared displacement (MSD) exhibits ballistic motion for several hundred beats before transitioning to diffusive behavior at the rotational diffusion timescale ($\approx$240\,\(T_b\)). Moreover, the propulsion speed versus curvature‐wave amplitude \(A\) follows the same trend as for a single Taylor‐line swimmer~\cite{khan2022effect}: it increases quadratically at low \(A\), reaches a maximum when the undulation amplitude approaches the radius of curvature (\(R_c = 1/C_0 \approx L/5\)), and then decreases at higher amplitudes.
}

Simulations of an \( FA \) with zero average curvature (\( FA_0 \)) show that although the overall shape deformation of the two flagella is non-reciprocal and breaks time symmetry, breaking time symmetry alone is a necessary but not sufficient condition for net propulsion. The \( FA_0 \) exhibited negligible random propulsion, with any minor displacement attributed to diffusion induced by thermal fluctuations.  This was further validated as its MSD asymptotically matched that of an inactive \( FA_{i0} \).
Furthermore, no significant enhancement in the diffusion coefficient of \( FA_0 \) was observed, consistent with the findings of Ref.~\cite{lauga2011enhanced}, which attribute this to the high beating frequency relative to the rotational diffusion relaxation time.

We determined the orientation correlation functions (OACFs) from the simulations and found that the OACFs of \( FA \) and \( FA_0 \) decay exponentially and match their inactive counterparts. This indicates that the beating activity does not influence rotational diffusion. We also simulated the dynamics of an asymmetric \( FA \), where the arms beat at different frequencies and reproduced the experimentally observed epitrochiod-like trajectory.

{\color{black}We studied hydrodynamic interactions between two \(FA\)s beating with a relative phase offset and found that bound pairs form for all \(\Delta\phi/2\pi\in[0,1]\). These pairs manifest as parallel-aligned bound pairs for \(\Delta\phi/2\pi\in[0,0.55]\cup[0.75,1]\), and as anti-parallel-aligned bound pairs for \(\Delta\phi/2\pi\in[0.3,0.7]\). Interestingly, the propulsion speed of anti-parallel-aligned bound pairs increases quadratically as \(|\Delta\phi/2\pi-0.5|\) grows. }

There are several directions in which this work can be expanded. Firstly, the role of hydrodynamic interactions may be explored when the two arms of the flagellar apparatus differ in phase, length, and mean curvature. Secondly, experimental observations show a V-shaped junction between the two flagella, where the angle can change over time. Future MPC simulations may incorporate this junction by introducing a different spring constant for the connecting springs and a potential that allows the angle to vary around a preferred value.  Given that the $FA$ forms pairs in various possible ways, examining the collective motion of multiple flagellar apparatuses and their cluster formation dynamics would be an interesting aspect.


In summary, we developed a computational model of a two armed swimmer which mimics the isolated flagellar apparatus of \textit{C. reinhardtii} and investigated various ascpects of its swimming dynamics and hydrodynamic interactions. These findings enhance our understanding of puller-type microswimmers and may provide valuable insights for the design and development of artificial microswimmers.
\section*{Author Contributions}
S.V.R. and A.G. jointly developed the conceptual framework, formulated the problem, and analyzed the MPC results. S.V.R. implemented the MPC code and ran the simulations, while A.B. performed the flow calculations using regularized Stokeslet method. S.V.R. drafted the initial manuscript, and all authors contributed to interpreting the results and revising the manuscript.
\section*{Conflicts of interest}
``There are no conflicts to declare''.
\section*{Data availability}
\textcolor{black}{The data supporting this article are included in the Supplementary Information. All code used to generate the results presented here is available at the following link: \textit{https://github.com/SaiVRamana/Multi-Particle-Collision-Dynamics-of-Flagellar-Apparatus}.}
\section*{Acknowledgements}
The authors acknowledge insightful discussions with Professor A. Pumir. We are also grateful to Dr. M. Y. Khan for his valuable support during the initial phase of this project. Lastly, we extend our thanks to the High Performance Computing Center at NYUAD for providing the computational resources and data storage for this work.
%
\bibliography{rsc} 
\bibliographystyle{rsc} 
\newpage
{\color{black}
	\appendix
\section{Appendix}	
\setcounter{equation}{0} 
\renewcommand{\theequation}{S\arabic{equation}}
\renewcommand\thefigure{S\arabic{figure}}
\setcounter{figure}{0}
\subsection{Method of regularized Stokeslets}
\label{Stokslets}
The length and velocity scales that characterize the locomotion of microswimmers are typically very small. As a result, the fluid flow generated by their motion is dominated almost entirely by viscous dissipation, as originally emphasized by Purcell~\cite{purcell2014life}. The full nonlinear Navier–Stokes equation of fluid flow reads:
\begin{equation}
	\rho_f\,\dot{\mathbf u} \;+\;\rho_f\,\mathbf u\!\cdot\!\nabla\mathbf u
	\;=\;-\,\nabla p \;+\;\mu\,\Delta\mathbf u\,,
\end{equation}
where, \( p \) is the pressure, \( \bm{u} \) is the fluid velocity, $\rho_f$ is the density and $\mu$ is the dynamic viscosity of the fluid. The inertial terms on the left‐hand side are usually negligible in the proximity of the microswimmer. The steady (classical) Reynolds number is defined as the ratio of the magnitude of the convective (inertial) term to the viscous diffusion term. Equivalently, it compares advective transport of momentum to its viscous diffusion and is most often written in the familiar form:
\[
\mathrm{Re}
\;=\;
\frac{\rho_f\,\lvert\mathbf u\!\cdot\!\nabla\mathbf u\rvert}
{\mu\,\lvert\Delta\mathbf u\rvert}
\;\sim\;
\frac{U\,L}{\nu},
\]
where \(U\) and \(L\) are characteristic speed and length scales of the swimmer and \(\nu = \mu/\rho_f\) is the kinematic viscosity. The unsteady (frequency‐based) Reynolds number compares the time‐dependent inertial term to viscous diffusion:
\[
\mathrm{Re}_{\omega}
\;=\;
\frac{\rho_f\,\lvert\partial_t\mathbf u\rvert}
{\mu\,\lvert\Delta\mathbf u\rvert}
\;\sim\;
\frac{\omega\,L^{2}}{\nu}.
\]

In our experiments, the swimming $FA$ has a contour length of approximately \( 20\,\mu\text{m} \), its beat frequency $\omega/2\pi$ is about 50 Hz and reaches a maximum swimming speed of about \( 100\,\mu\text{m/s} \). Using the kinematic viscosity of water, \( \nu \approx 10^{-6}\,\text{m}^2/\text{s} \), the steady and unsteady Reynolds numbers are estimated as:
\begin{align}
\text{Re} &= \frac{UL}{\nu} \approx \frac{(100 \times 10^{-6})(20 \times 10^{-6})}{10^{-6}} \sim 2 \times 10^{-3},\\
\text{Re}_\omega&=\frac{\omega\,L^{2}}{\nu}\approx \frac{(2\pi \times 50)(20 \times 10^{-6})^2}{10^{-6}} \sim 10^{-1}
\end{align}
indicating that the flow is well within the low-Reynolds-number regime. Nevertheless, inertial effects become important at large distances~\cite{klindt2015flagellar,wei2019zero}.  In a force‐dipole flow field,
\[
\rho_f\,\dot{\mathbf u}\sim \rho_f\,\omega\,r^{-2}, 
\quad 
\mu\,\Delta\mathbf u\sim r^{-4},
\quad 
\rho_f\,\mathbf u\!\cdot\!\nabla\mathbf u\sim r^{-5},
\]
so that unsteady acceleration dominates beyond the characteristic length $
\delta \;=\;\sqrt{2\nu/\omega}\sim100\mu\text{m}\, $~\cite{landau1991hydrodynamik}.

Since our analysis is confined to distances less than $\delta$ from the swimmer, we adopt the zero–Reynolds–number approximation. In this regime, the fluid motion is governed by the Stokes equations:
\begin{align}
	\nabla \cdot \bm{\sigma} &= -\nabla p + \mu \Delta \bm{u} = 0, \label{eq:stokes_momentum} \\
	\nabla \cdot \bm{u} &= 0, \label{eq:stokes_incompress}
\end{align}
where \( \bm{\sigma} = -p \bm{I} + 2\mu \bm{E} \) is the Newtonian fluid stress tensor. The identity tensor is denoted by \( \bm{I} \), and the symmetric rate-of-strain tensor is given by \( \bm{E} = (\nabla \bm{u} + \nabla \bm{u}^{\mathrm{T}})/2 \). 
Applying Green’s theorem to the Stokes Eqs.~\eqref{eq:stokes_momentum}-~\eqref{eq:stokes_incompress} yields an integral representation of the fluid velocity that depends solely on the stress and velocity distributions along the immersed boundaries—namely, the surface of the swimming body~\cite{pozrikidis1992boundary}. The fluid velocity is assumed to decay to zero in the far field, and appropriate boundary conditions are imposed based on the geometry of the swimming body. In scenarios with a solid boundary—such as our experiments, where the \(FA\) swims in close proximity to a glass substrate modeled as an infinite plane at \(z = z_w\)—we enforce the no‐slip boundary condition, $\mathbf{u}(z_w) \;=\; \mathbf{0}.$

To impose this no-slip condition on the wall, we employ the method of images in combination with regularized Stokeslets, as introduced by Ainley \emph{et al.}~\cite{ainley2008method}. This approach builds on the original formulation developed by Cortez~\cite{cortez2001method}. In this framework, the fluid flow generated by a swimmer is modeled using \( N \) regularized point forces $\mathcal{F}=(\bm{f}_1,...,\bm{f}_k,...,\bm{f}_N)^T$  (a $3N\times 1$ vector), applied at locations 
\[
\bm{x}_{k,0} = (x_k, y_k, z_w+d_k)
\]
in the fluid domain (i.e. $d_k>0$) on the surface of the swimmer. Here, $d_k$
is simply the distance from the $k$-th point force to the wall. To satisfy the no-slip boundary condition at the wall, the method also incorporates contributions from corresponding image points,
\[
\bm{x}_{k,\text{im}} = (x_k, y_k, z_w-d_k).
\]

The regularized Stokeslet method avoids the singularities associated with classical Stokeslets by replacing the point force (Dirac delta function) with a smooth, localized distribution defined by a shape function—commonly referred to as a \textit{blob}. The spread of this distribution is controlled by a regularization parameter \( \delta \), which determines the size of the region over which the force is applied. As a result, for \( \delta > 0 \), the expressions for the regularized Stokeslet remain finite everywhere and converge to the classical Stokeslet in regions away from the singularity. For a given point \( \bm{x} \) in the fluid domain, or on the surface of the swimmer or the wall, we define: \( \bm{x}_k^* = \bm{x} - \bm{x}_{k,0} \), 
and 
\( \bm{x}_k = \bm{x} - \bm{x}_{k,\text{im}}  \). Then if we use the \textit{blob} $\psi(r) = \frac{15 \delta^4}{8\pi (r^2 + \delta^2)^{7/2}}$, which spreads the singular effect of a point force to a small finite area, the velocity at \( \bm{x} \) is given by:

{\small
	\begin{align}
		\bm{u}(\bm{x})& =\mathcal{M}\mathcal{F}=\sum_{k=1}^{N} \Big[
		\bm{f}_k H_1(|\bm{x}_k^*|) 
		+ (\bm{f}_k \cdot \bm{x}_k^*) \bm{x}_k^* H_2(|\bm{x}_k^*|) 	\Big]\notag \\
		&\quad - \Big[\bm{f}_k H_1(|\bm{x}_k|) 
		+ (\bm{f}_k \cdot \bm{x}_k) \bm{x}_k H_2(|\bm{x}_k|) \Big]
		- h_k^2 \Big[ \bm{g}_k D_1(|\bm{x}_k|) \notag \\
		&\quad + (\bm{g}_k \cdot \bm{x}_k) \bm{x}_k D_2(|\bm{x}_k|) \Big]
		+2 h_k \Big[ \frac{H_1'(|\bm{x}_k|)}{|\bm{x}_k|}+H_2(|\bm{x}_k|)\Big](\bm{L}_k \times \bm{x}_k) \notag \\
		&\quad + 2 h_k \Bigg[
		(\bm{g}_k \cdot \hat{\bm{z}}) \bm{x}_k H_2(|\bm{x}_k|) 
		+ (\bm{x}_k \cdot \hat{\bm{z}}) \bm{g}_k H_2(|\bm{x}_k|) \notag \\
		&\qquad + (\bm{g}_k \cdot \bm{x}_k) \hat{\bm{z}} \frac{H_1'(|\bm{x}_k|)}{|\bm{x}_k|}+ (\bm{x}_k \cdot \hat{\bm{z}})(\bm{g}_k \cdot \bm{x}_k) \bm{x}_k \frac{H_2'(|\bm{x}_k|)}{|\bm{x}_k|} 
		\Bigg],
		\label{u_cortez}
	\end{align}
}
where $\mathcal{M}$ is a $3N\times3N$ matrix and the dipole strengths and rotlet strengths are given by:
\[
\bm{g}_k = 2(\bm{f}_k \cdot \hat{\bm{z}})\hat{\bm{z}} - \bm{f}_k, 
\quad 
\bm{L}_k = \bm{f}_k \times \hat{\bm{z}},
\]
and the regularized kernel functions are:
\begin{align}
	H_1(r) &= \frac{1}{8\pi (r^2 + \delta^2)^{1/2}} + \frac{\delta^2}{8\pi (r^2 + \delta^2)^{3/2}}, \notag \\
	H_2(r) &= \frac{1}{8\pi (r^2 + \delta^2)^{3/2}}, \notag \\
	D_1(r) &= \frac{1}{4\pi (r^2 + \delta^2)^{3/2}} - \frac{3\delta^2}{4\pi (r^2 + \delta^2)^{5/2}}, \notag \\
	D_2(r) &= -\frac{3}{4\pi (r^2 + \delta^2)^{5/2}}. \notag
\end{align}
\begin{figure*}[t!]
	\centering
		\includegraphics[width=1.0\columnwidth]{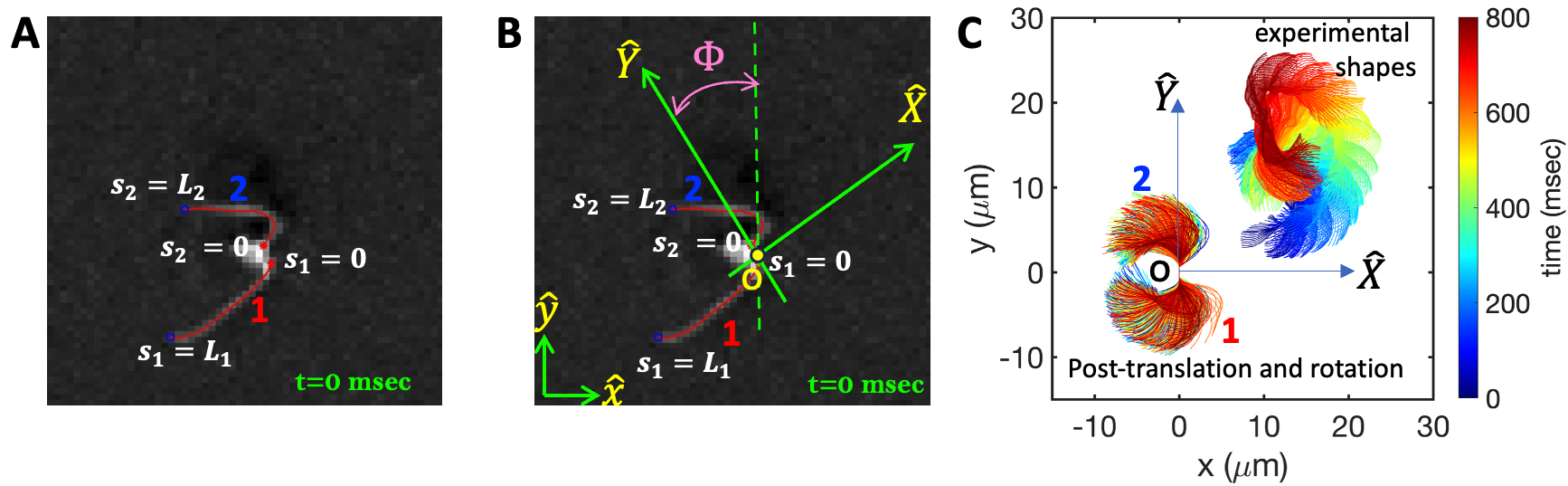}
	\includegraphics[width=1.0\columnwidth]{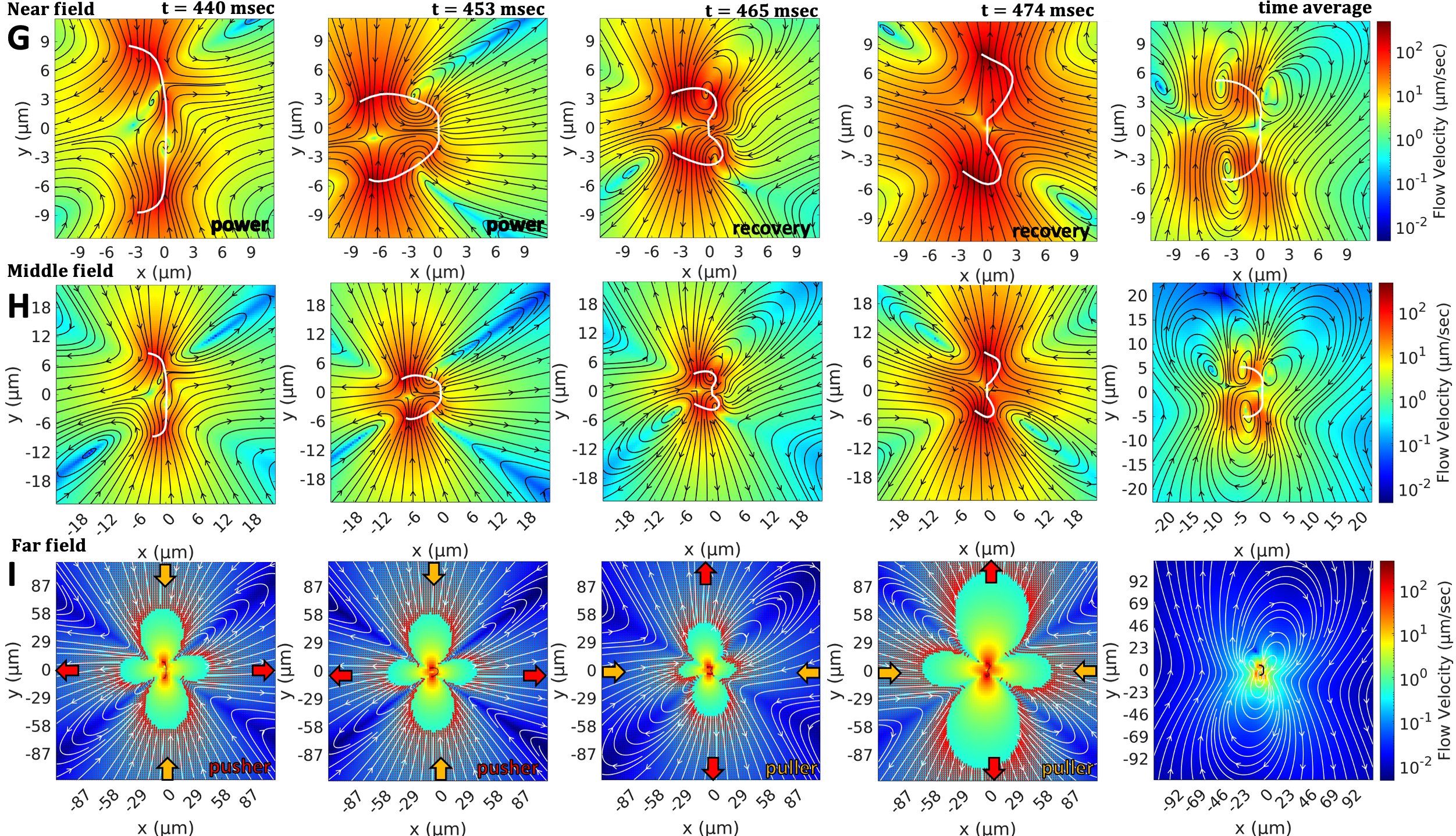}
	\caption{ \textcolor{black}{A-C) We computed the flow field throughout the 3D domain using our experimental \(FA\) shapes swimming far away from a substrate (at \(1000\) pixels \(\approx455~\mu\)m above the surface), employing the method of images for the regularized Stokeslet developed by Ainley \emph{et al.}~\cite{ainley2008method}. Analysis of the far-field flow reveals that the \(FA\) alternates between puller and pusher modes, averaging to a behavior characteristic of a nearly neutral swimmer.} }
	\label{fig:FA_EXP_Z0_Minus1000}
\end{figure*}

 To compute the localized forces \(\bm{f}_k\) at parts along the swimmer’s surface, we evaluate Eq.~\eqref{u_cortez} at \(\bm{x} = \bm{x}_{k,0}\). By enforcing the no‐slip condition, the fluid velocity at the surface, \(\bm{u}(\bm{x}_{k,0})\), must match the swimmer’s measured surface velocity \(\bm{U}_{\text{swimmer}}(\bm{x}_{k,0})\). We then obtain the force vector \(\mathcal{F}\) by inverting the  matrix \(\mathcal{M}\):
 \begin{equation}
 \mathcal{F} \;=\; \mathcal{M}^{-1}\,\bm{u}(\bm{x}_{k,0})
 \;=\;
 \mathcal{M}^{-1}\,\bm{U}_{\text{swimmer}}(\bm{x}_{k,0})\,.
 \label{cortez:invert}
\end{equation}
 Substituting these forces back into Eq.~\eqref{u_cortez} allows us to compute the fluid velocity field throughout the domain.

Using the method outlined above, we now summarize the steps taken to compute the three-dimensional flow field from experimentally obtained \(FA\) shapes. Because an \(FA\) swimming in a fluid produces low--Reynolds--number flows, the net fluid force \(\mathbf{F}^{FA}\) and torque \(\mathbf{T}^{FA}\) on the \(FA\) both vanish. In this analysis, we neglect the basal body’s contribution and consider only the two flagella; thus, \(\mathbf{F}^{FA}\) consists solely of the hydrodynamic forces on these flagella:
\begin{equation}
\mathbf{F}^{FA}
=
\int_{0}^{L_{1}} \mathbf{F}_{1}(s,t)\,\mathrm{d}s
+
\int_{0}^{L_{2}} \mathbf{F}_{2}(s',t)\,\mathrm{d}s'
\;=\;0\,,
\label{Force}
\end{equation}
where \(L_{1}\) and \(L_{2}\) are the contour lengths of the first and second flagella, respectively.  
Similarly, the total hydrodynamic torque acting on the $FA$ must vanish:
\begin{equation}
\mathbf{T}^{FA}
=
\int_{0}^{L_{1}} \bigl[\mathbf{r}_{1}(s,t)\times \mathbf{F}_{1}(s,t)\bigr]\,\mathrm{d}s
+
\int_{0}^{L_{2}} \bigl[\mathbf{r}_{2}(s,t)\times \mathbf{F}_{2}(s,t)\bigr]\,\mathrm{d}s
\;=\;0\,.
\label{Torque}
\end{equation}
The $FA$, having two deformable flagella, may at any instant be considered as a rigid body with unknown translational velocity \(\mathbf{U}(t)\) and rotational velocity \(\boldsymbol{\Omega}(t)\).  
The force \(\mathbf{F}^{FA}\) and torque \(\mathbf{T}^{FA}\) exerted by the fluid on the $FA$—given by Eqs.~\ref{Force}-\ref{Torque}—can be decomposed into a propulsive part (due to the relative deformation of both flagella in a body‐fixed frame) and a drag part:
\begin{equation}
\begin{pmatrix}
	\mathbf{F}^{FA} \\[6pt]
	\mathbf{T}^{FA}
\end{pmatrix}
=
\begin{pmatrix}
	\mathbf{F}^{\mathrm{prop}} \\[6pt]
	\mathbf{T}^{\mathrm{prop}}
\end{pmatrix}
\;-\;
\mathbf{A}
\begin{pmatrix}
	\mathbf{U} \\[3pt]
	\boldsymbol{\Omega}
\end{pmatrix}=0\,.
\label{Matrix}
\end{equation}
Here, \(\mathbf{A}\) is a \(6\times 6\) drag‐coefficient matrix—symmetric, invertible, and determined solely by the \(FA\)’s shape. At each time \(t\), we compute \(\mathbf{A}\) by evaluating two cases:

1. Translating, Non‐Rotating \(FA\):  Set the translational velocities to:
\[
\mathbf{U} = (1,0,0),\quad (0,1,0),\quad (0,0,1),
\]
corresponding to translations by unity along \(\hat{x}\), \(\hat{y}\), or \(\hat{z}\), respectively.

2. Rotating, Non‐Translating \(FA\):  
Set the rotational velocities around each axis to:
\[
\boldsymbol{\Omega} = (1,0,0),\quad (0,1,0),\quad (0,0,1).
\]
For each rotation, the translational velocity is given by:
\[
\mathbf{U} = \mathbf{r} \times \boldsymbol{\Omega}, 
\quad \text{with} ~~\;\mathbf{r} = (\mathbf{x},\mathbf{y},\mathbf{z}).
\]
Thus, for 
\(\boldsymbol{\Omega} = (1,0,0)\), 
\(\mathbf{U} = (0,\,\mathbf{z},\,-\mathbf{y})\);  
for \(\boldsymbol{\Omega} = (0,1,0)\), 
\(\mathbf{U} = (-\mathbf{z},\,0,\,\mathbf{x})\);  
and for \(\boldsymbol{\Omega} = (0,0,1)\), 
\(\mathbf{U} = (\mathbf{x},\,-\mathbf{y},\,0)\).  

For each of these six configurations, we compute the localized forces exerted by the \(FA\) on the fluid using Eq.~\ref{cortez:invert}, with given \(z_w\) and \(\delta\). The resulting force–torque pairs are then substituted into Eq.~\ref{Matrix} to populate the corresponding columns of \(\mathbf{A}\).

In our analysis, we treat the two flagella as a single composite swimmer. Neglecting the basal body, we connect the point \(s_{1}=0\) on the first flagellum to the point \(s_{2}=0\) on the second flagellum with a straight segment, thus forming one combined swimmer. We then introduce a swimmer‐fixed coordinate system whose origin lies at the midpoint of this segment. The vector from
 \(s_{1}=0\) to \(s_{2}=0\) defines the \(\hat{Y}\)‐axis, the perpendicular direction defines the \(\hat{X}\)‐axis, and we assume \(\hat{z}\) is parallel to \(\hat{Z}\). Let \(\Phi(t)\) denote the angle between \(\hat{y}\) and \(\hat{Y}\) (see Fig.~\ref{fig:FA_EXP_Z0_Minus1000}A–B). 
First, for each time point, we translate and rotate the \(FA\) shape by \(\Phi(t)\) so that the origin \(O\) is at \((0,0)\) and the \(\hat{Y}\)‐axis aligns with \(\hat{y}\) (see Fig.~\ref{fig:FA_EXP_Z0_Minus1000}C). This step, which gives us $FA$ shapes in the body frame i.e. $\mathbf{r}_{\text{body–frame}}(s,t)=(\mathbf{x}_{\text{body}}(s,t),\mathbf{y}_{\text{body}}(s,t),\mathbf{z}_{\text{body}}(s,t),1)$, removes all orientation information except at \(t=0\). Second, using these shapes, we compute the deformation velocities of the \(FA\) between time points \(t\) and \(t+dt\) and use Eq.~\ref{cortez:invert} to determine localized forces exerted by the swimmer on the fluid. Third, we use these localized forces to  calculate torque as $\mathbf{r}_{\text{body–frame}}(s,t)\times\mathcal{F}$ and then insert  them into Eq.~\ref{Matrix} to calculate the required global translation $\mathbf{U}=(U_x,U_y,U_z)$ and rotation $\boldsymbol{\Omega}=(\Omega_x,\Omega_y,\Omega_z)$ in 3D that enforce a zero total force and torque on the swimmer. Having calculated \(\mathbf{U}\) and \(\boldsymbol{\Omega}\), the incremental rotation–translation matrix is given by Rodrigues’ formula augmented with a translational shift:

\[
d\mathbf{R}(\mathbf{u},\theta) \;=\;
\begin{pmatrix}
	\cos\theta + w_x^2\bigl(1 - \cos\theta\bigr) 
	& w_x\,w_y\bigl(1 - \cos\theta\bigr) \;-\; w_z\,\sin\theta 
	& w_x\,w_z\bigl(1 - \cos\theta\bigr) \;+\; w_y\,\sin\theta&U_x \;dt \\[8pt]
	w_y\,w_x\bigl(1 - \cos\theta\bigr) \;+\; w_z\,\sin\theta 
	& \cos\theta + w_y^2\bigl(1 - \cos\theta\bigr) 
	& w_y\,w_z\bigl(1 - \cos\theta\bigr) \;-\; w_x\,\sin\theta&U_y \;dt \\[8pt]
	w_z\,w_x\bigl(1 - \cos\theta\bigr) \;-\; w_y\,\sin\theta 
	& w_z\,w_y\bigl(1 - \cos\theta\bigr) \;+\; w_x\,\sin\theta 
	& \cos\theta + w_z^2\bigl(1 - \cos\theta\bigr)&U_z \;dt \\[8pt]
	0&0&0&1
\end{pmatrix}.
\]

This matrix rotates the \(FA\) by an angle \(\theta = \lVert \boldsymbol{\Omega} \rVert \,dt\) around the unit axis $\mathbf{w}=(w_x,w_y,w_z)^T=(\Omega_x/\lVert \boldsymbol{\Omega}\rVert,\Omega_y/\lVert \boldsymbol{\Omega}\rVert,\Omega_z/\lVert \boldsymbol{\Omega}\rVert)^T$ and translates by $(U_x \;dt, U_y \;dt, U_z \;dt)$.  Now we use $d\mathbf{R}(\mathbf{u},\theta)$ to update the full rotation matrix via
\[
\mathbf{R}(t+dt) \;=\; \mathbf{R}(t)\,d\mathbf{R}(t,\mathbf{u},\theta),
\]
taking \(\mathbf{R}(t = 0)\) to be the identity matrix. Having obtained \(\mathbf{R}(t)\) at each time \(t\), we recover the laboratory‐frame configuration of the $FA$ from its body‐fixed shape by
\[
\mathbf{r}_{\text{lab–frame}}(s,t)
\;=\;
\mathbf{R}(t)\,\mathbf{r}_{\text{body–frame}}(s,t).
\]
These reconstructed 3D shapes can be compared with experiment. Since our data provide only 2D projections of the \(FA\) and the exact height \(z_w\) above the glass substrate is unknown, we can adjust \(z_w\) to achieve good agreement between reconstructed and experimental shapes. However, our primary goal is to compute the flow profile throughout the 3D domain for a fixed \(z_w\). To do so, we use the laboratory‐frame shapes \(\mathbf{r}_{\text{lab–frame}}(s,t)\) obtained above to determine the \(FA\) deformation velocity in the laboratory frame, which—by the no‐slip condition—equals the fluid velocity at the \(FA\) surface. Substituting this velocity into Eq.~\ref{cortez:invert} yields the localized forces  exerted by the swimmer on the fluid; these, in turn, are used in Eq.~\ref{u_cortez} to compute the velocity field throughout the 3D domain. Results for two values of \(z_w\) are shown in Fig.~\ref{fig:FA_EXP}C–I (\(z_w =20~\text{pixels}\approx 9\,\mu\mathrm{m}\), close to the substrate) and Fig.~\ref{fig:FA_EXP_Z0_Minus1000}D–F (\(z_w =1000~\text{pixels}\approx 455\,\mu\mathrm{m}\), far from the substrate).

\subsection{Artificial clustering}
\label{clustering}
In simulations of squirmer collective behaviour using the MPC method, Theers \emph{et al.}~\cite{theers2018clustering} found that at low average fluid density (\(\rho_{0}\approx 10\)), the MPC fluid was expelled from collision cells containing squirmers, resulting in segregation into high‐ and low‐density regions. In areas with squirmer clusters, the fluid became very dilute, and some cells even appeared empty. Conversely, regions without squirmers became highly dense. This artifact led to spurious squirmer clustering, which Theers \emph{et al.} attributed to the MPC fluid’s compressibility. Since our simulations also use \(\rho_{0}\approx 10\), we investigate below whether a similar effect arises for \(FA\)s.
\begin{figure}[!b]
	\centering
	\includegraphics[width=1.0\linewidth]{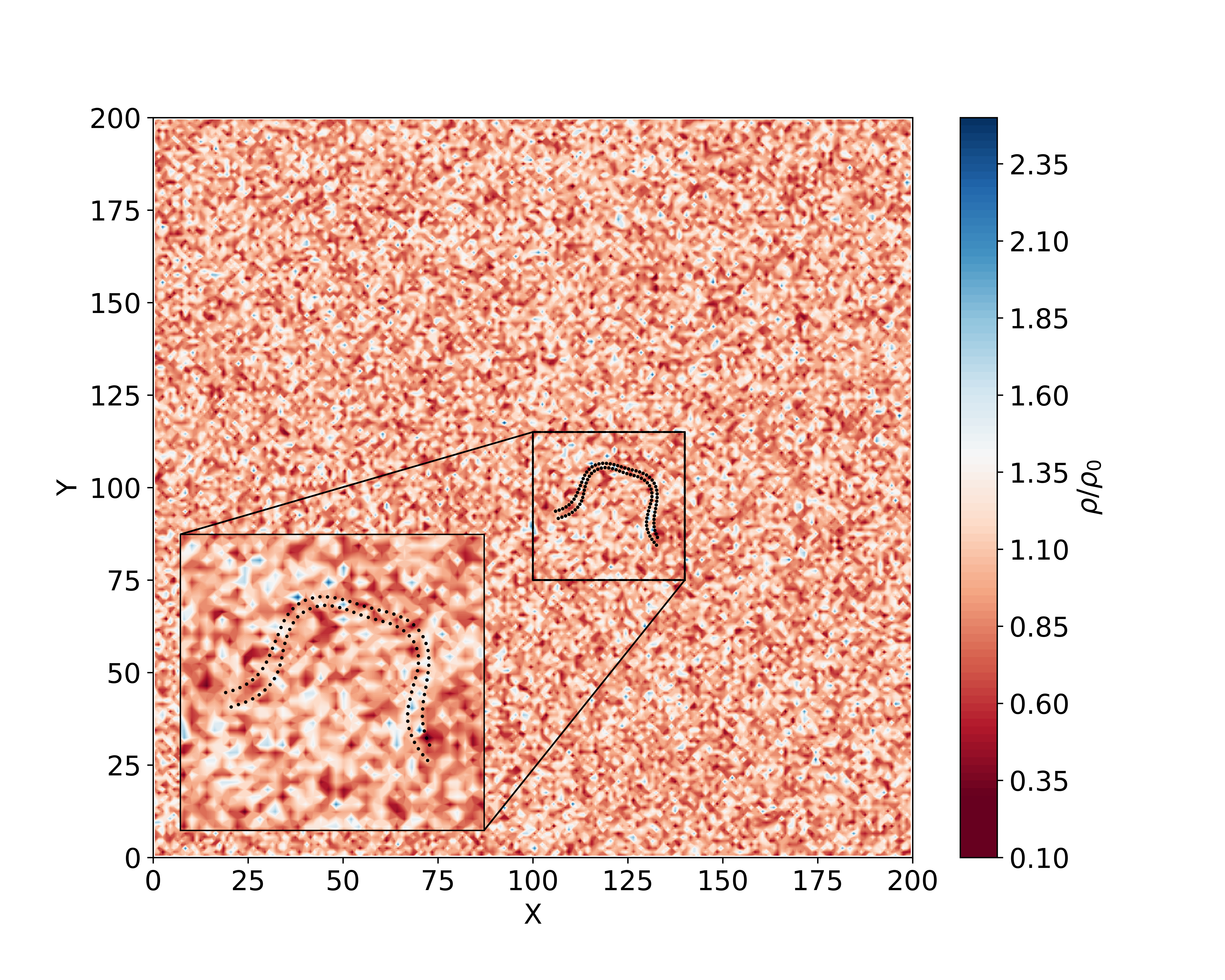}
	\caption{\textcolor{black}{MPC fluid density at a single time point in a simulation box containing a co‐moving pair of \(FA\)s. The color bar shows \(\rho/\rho_{0}\) with \(\rho_{0} = 10\). Regions with \(\rho/\rho_{0} < 0.3\) appear dark red, and those with \(\rho/\rho_{0} > 2.5\) appear dark blue. Although small‐scale density fluctuations of order \(a\) are visible, there is no large‐scale segregation into high‐ and low‐density domains. The inset provides a zoomed‐in view of the \(FA\) and its immediate fluid surroundings. }}
	\label{fig:nT2FliudDen}
\end{figure}

Figure~\ref{fig:nT2FliudDen} displays a heatmap of the MPC fluid density at an instant in a simulation containing a co‐moving pair of \(FA\)s with zero phase difference—chosen because this configuration minimizes the gap between them. Although no cells are entirely empty, the figure reveals substantial density fluctuations: small granular regions of higher and lower density are distributed uniformly throughout the box. The inset provides a zoomed‐in view of the \(FA\) pair and its surrounding fluid, showing that the fluctuations in the space between and around the swimmers are no different from those elsewhere. Therefore, we conclude that the attraction between the \(FA\)s is driven by hydrodynamic interactions rather than by expulsion of fluid from the gap. For comparison, we also simulated the same co‐moving \(FA\) pair with \(\rho_{0} = 80\). Figure~\ref{fig:nT2FluidDen80} shows the corresponding density heatmap. Although density fluctuations are still present, their magnitude is considerably reduced compared to the \(\rho_{0}=10\) case.
\begin{figure}[!t]
    \centering
    \includegraphics[width=1.0\linewidth]{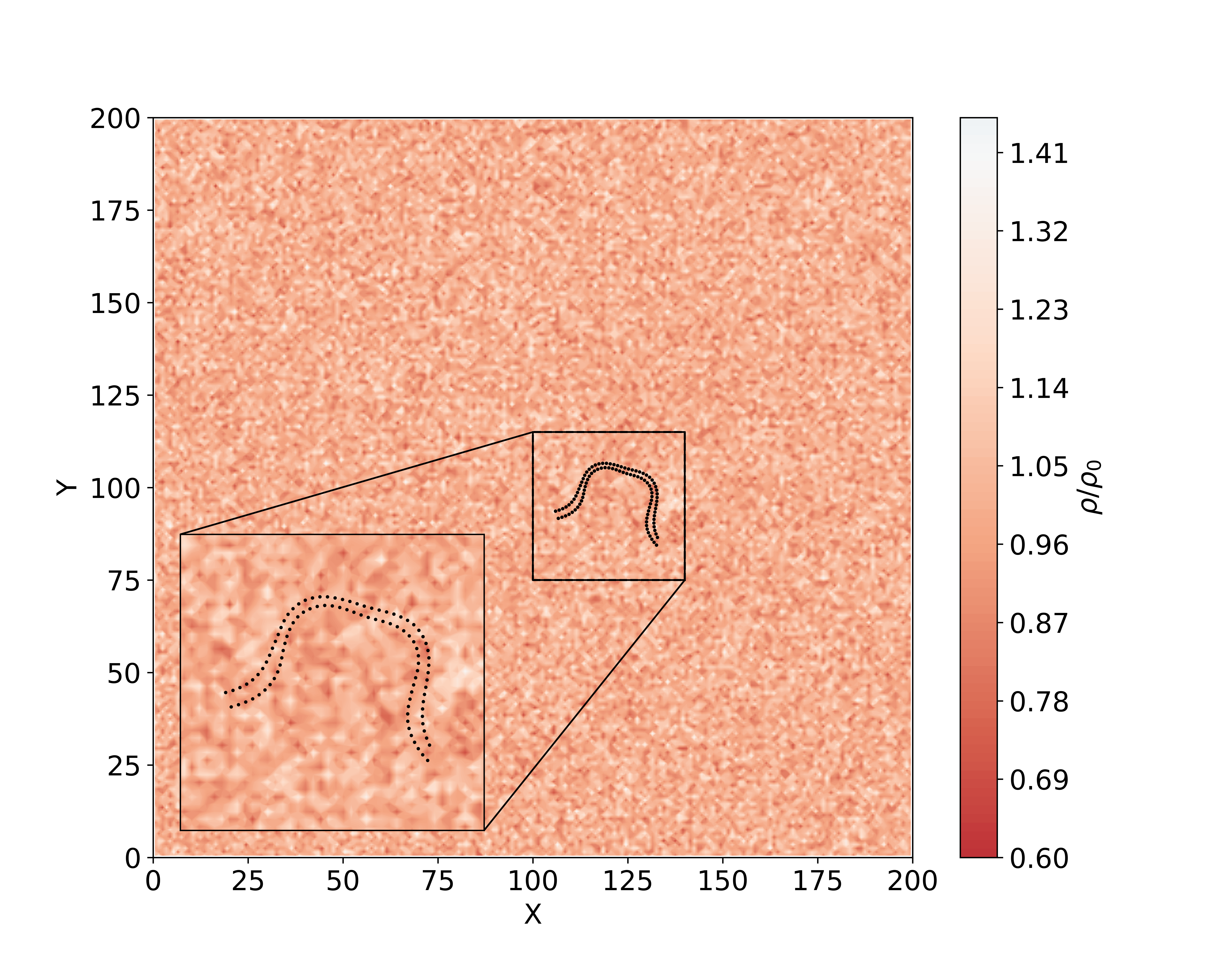}
    \caption{\textcolor{black}{Heatmap of the simulation box containing \(N=2\) \(FA\)s at \(\rho_{0}=80\), shown at a single time point (format as in Fig.~\ref{fig:nT2FliudDen}). Compared to the \(\rho_{0}=10\) case, the density fluctuations are noticeably reduced, although they remain significant.}
    }
    \label{fig:nT2FluidDen80}
\end{figure}

\begin{figure}[!t]
    \centering
    \includegraphics[width=1.0\linewidth]{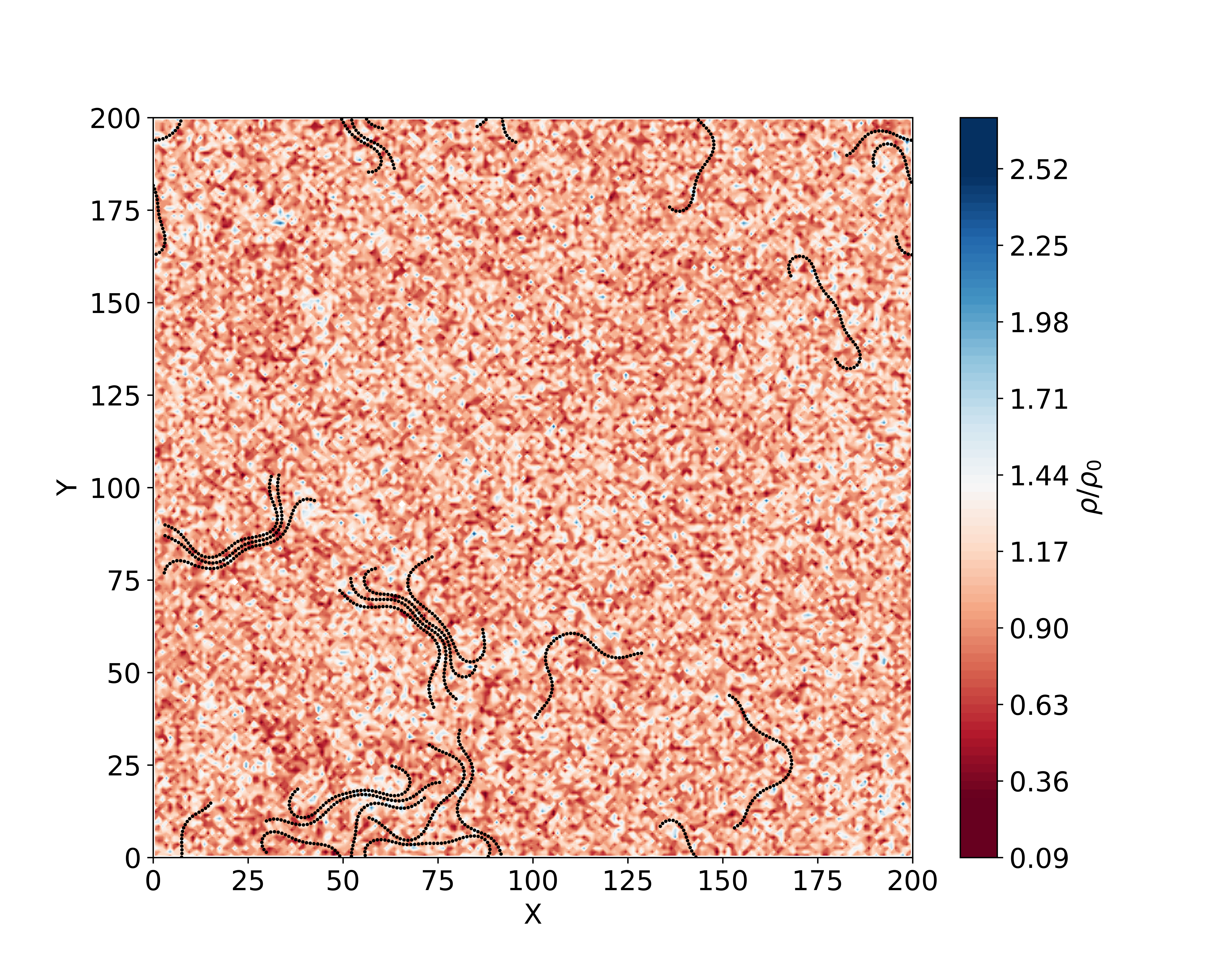}
    \caption{\textcolor{black}{Heatmap of the simulation box containing \(N=20\) \(FA\)s. The color bar shows \(\rho/\rho_{0}\) with \(\rho_{0} = 10\). No low‐density regions form around the \(FA\) clusters.}}
    \label{fig:nT20FluidDen}
\end{figure}
\begin{figure}[!htbp]
    \centering
    \includegraphics[width=1.0\linewidth]{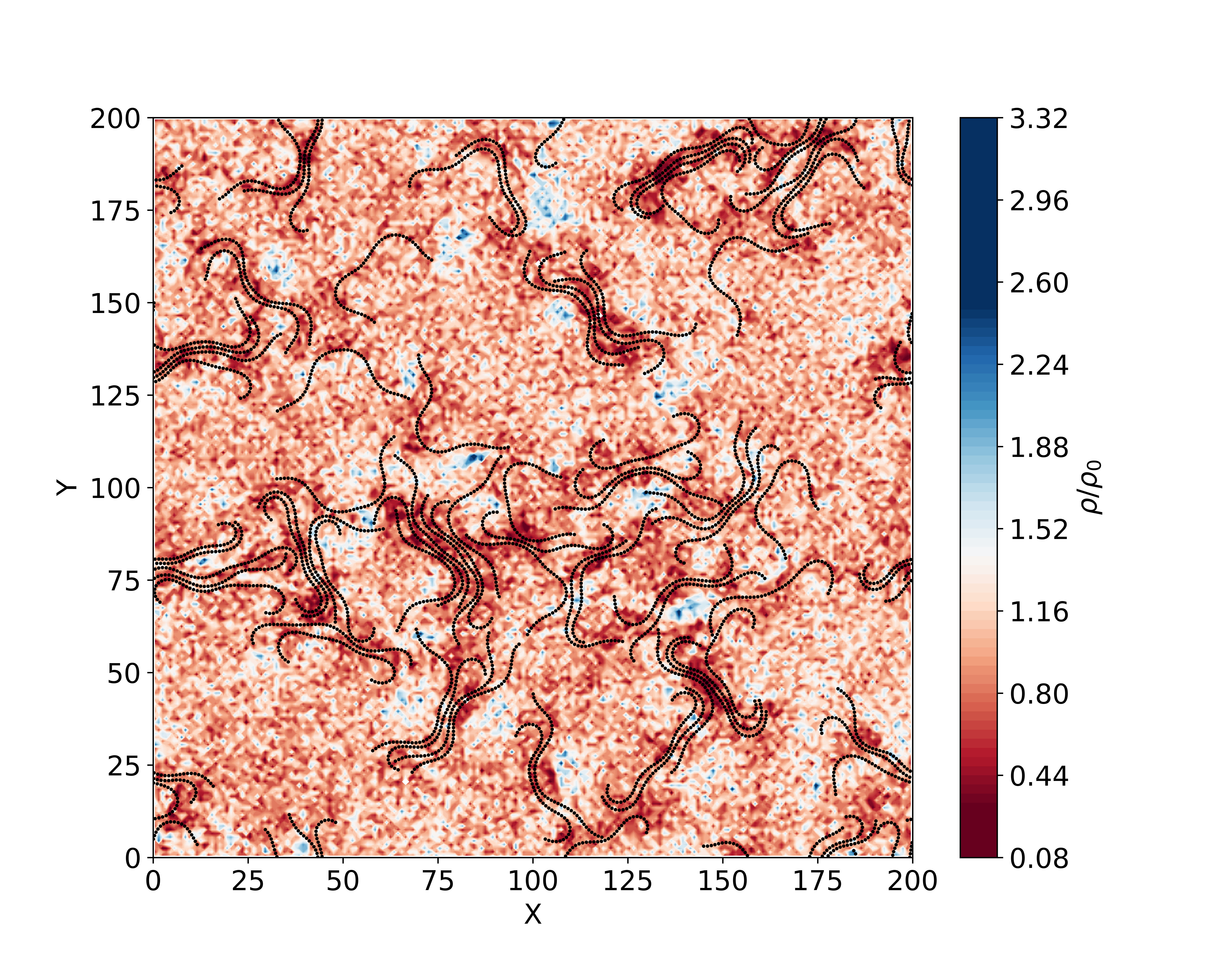}
    \caption{\textcolor{black}{Heatmap of the simulation box containing \(N=80\) \(FA\)s (same format as Fig.~\ref{fig:nT2FliudDen}). Large, low‐density regions form around clusters of \(FA\)s, while high‐density domains appear where no \(FA\)s are present. The probability of empty cells is an order of magnitude higher for \(N=80\) than for \(N=2\).}}
    \label{fig:nT80FluidDen}
\end{figure}

We hypothesized that the artifact observed by Theers \emph{et al.}~\cite{theers2018clustering} arises only when large clusters of active particles form. To test this, we ran simulations with \(N=20\) and \(N=80\) \(FA\)s. In each case, \(FA\)s were placed randomly in the box, and their beating phases were chosen uniformly at random. The system was evolved for several hundred beat periods until clusters appeared. Figure~\ref{fig:nT20FluidDen} shows the fluid density at a single time point for \(N=20\). Here, we observe one cluster of size 3, one of size 4, a larger cluster of size 7, and several isolated \(FA\)s. The density fluctuations remain relatively uniform throughout the box, including around the clusters, with no large low-density or high-density domains. By contrast, Fig.~\ref{fig:nT80FluidDen} (for \(N=80\)) exhibits pronounced segregation: large, low-density regions appear where clusters of \(FA\)s have formed, and high-density domains arise where no \(FA\)s are present. The density fluctuations are unevenly distributed across the box, closely mirroring the behavior reported by Theers \emph{et al.}~\cite{theers2018clustering}.

For \(N=2\) and \(N=20\), we occasionally observed a small number of empty cells (fewer than five) in a given snapshot, but these vanished by the next time step. In contrast, for \(N=80\), many snapshots contained empty cells that persisted over multiple time steps. Moreover, as clusters formed and grew, the fluid continuously segregated into distinct high‐ and low‐density regions whose domain sizes increased over time. Figure~\ref{fig:cdf} shows the cumulative distribution function (CDF), \(P(\rho/\rho_{0} < x)\), obtained by compiling density histograms over 20 consecutive time steps. In all cases, \(P(\rho/\rho_{0} < x)\) is negligible for \(x < 0.3\). However, at very low densities, the CDF for \(N=80\) indicates a probability of order \(10^{-4}\) for \(x < 0\), whereas for \(N=2\) it is of order \(10^{-5}\). This ten‐fold increase explains the more frequent and persistent empty cells when \(N=80\).

Theers \emph{et al.}~\cite{theers2018clustering} introduced a dimensionless “pump number,” \(Pu\), to predict whether low‐density regions will form at a given MPC fluid density \(\rho_{0}\). It is defined as the ratio of (i) the time for an MPC fluid particle to diffuse across the diameter \(\sigma\) of an active swimmer to (ii) the time for the swimmer to advect that particle via its surface activity. Mathematically,
\[
Pu \;=\; \frac{\sigma\,\langle S\rangle}{4\,D_{f}},
\]
where \(\langle S\rangle\) is the swimmer’s speed and \(D_{f}\) is the MPC fluid’s self‐diffusion coefficient. The factor of 4 in the denominator accounts for our two‐dimensional simulations.

The condition \(Pu \ll 1\) ensures that density inhomogeneities do not occur. Although \(Pu\) was originally defined for a spherical squirmer, we adapt it here for the \(FA\). In our model, each bead of the \(FA\) acts as an active particle, so we take \(\sigma \approx 1\). The time‐averaged speed is \(\langle S \rangle \approx 0.07\,L/T_{b}\). Noguchi and Gompper~\cite{noguchi2008transport} derived an analytical expression for the self‐diffusion coefficient \(D_{f}\) of an MPC fluid coupled to an Andersen thermostat with angular‐momentum conservation (MPC‐AT+a), namely
\begin{equation}
    D_f = \frac{k_BT \delta t}{m}\left(\frac{\rho_0/A}{\rho_0 -(d+1)/2}-1/2\right) \label{Df}
\end{equation}
In the equation above, we use \(A = 1\) for the Andersen thermostat, \(K_{B}T = 1\), and \(m = 1\) in MPC units. The spatial dimension is \(d = 2\), the time step is \(\delta t = 0.025\), and the average fluid density is \(\rho_{0} = 10\). Substituting these values yields
\[
D_{f} = 0.017 
\qquad\text{and}\qquad
Pu \approx 0.43.
\]
Since \(Pu < 1\), the likelihood of empty cells forming is low.

Therefore, when the MPC fluid has \(\rho_{0} = 10\), unphysical segregation into low‐density regions only occurs in the presence of large clusters of \(FA\)s. For smaller systems with \(N < O(10)\) swimmers, using \(\rho_{0} = 10\) produces physically reliable results.
\begin{figure}[!t]
    \centering
    \includegraphics[width=0.8\linewidth]{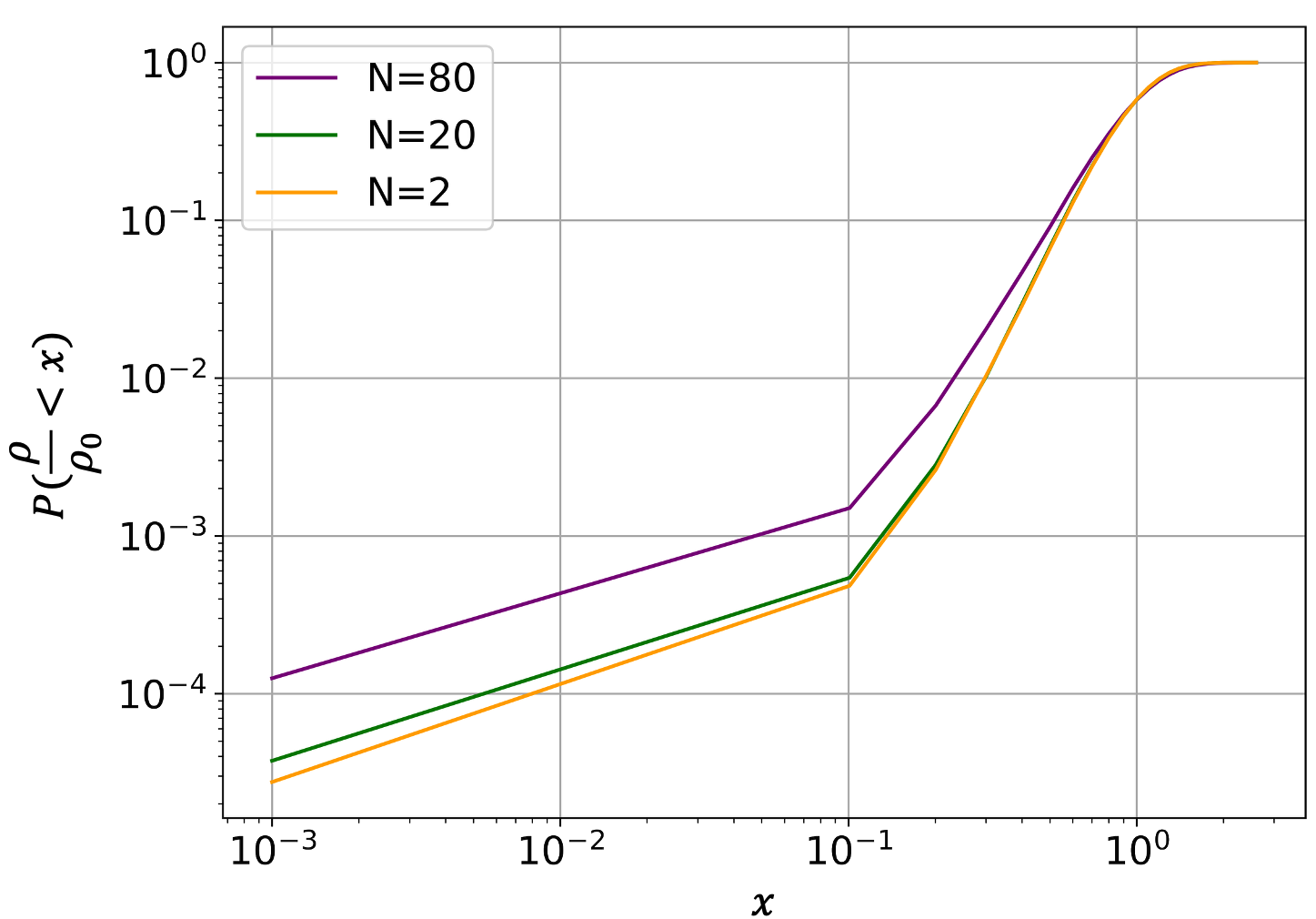}
    \caption{{\color{black}Cumulative distribution functions \(P\bigl(\rho/\rho_0 < x\bigr)\) versus \(x\) for the particle density. Note that \(P(\rho/\rho_0 < x)\) is negligible for \(x < 0.3\). A log–log scale is used to highlight variations in the CDF at low densities. All \(x\) values have been shifted by \(10^{-3}\).
    } }
    \label{fig:cdf}
\end{figure}

\subsection{Finite Size effects}
{\textcolor{black}{In this study, one might expect finite‐size effects because hydrodynamic interactions decay slowly (in low‐Reynolds‐number Stokes  flow, the Oseen tensor’s Green’s function exhibits a \(\sim 1/r\) decay). However, Goldstein \emph{et al.}~\cite{drescher2011fluid} demonstrated that thermal noise and the swimmers’ intrinsic stochasticity effectively screen these long‐range flows, limiting hydrodynamic interactions to lengths on the order of the swimmer size. As a result, the remaining short‐range interactions drive collective behavior in bacterial suspensions, much like in other driven systems—such as granular media, bird flocks, and vehicular traffic~\cite{krapivsky2010kinetic,ben1994kinetics,sai2022stationary,venkata2020traffic}.}
\begin{figure}[!b]
    \centering
    \includegraphics[width=0.8\linewidth]{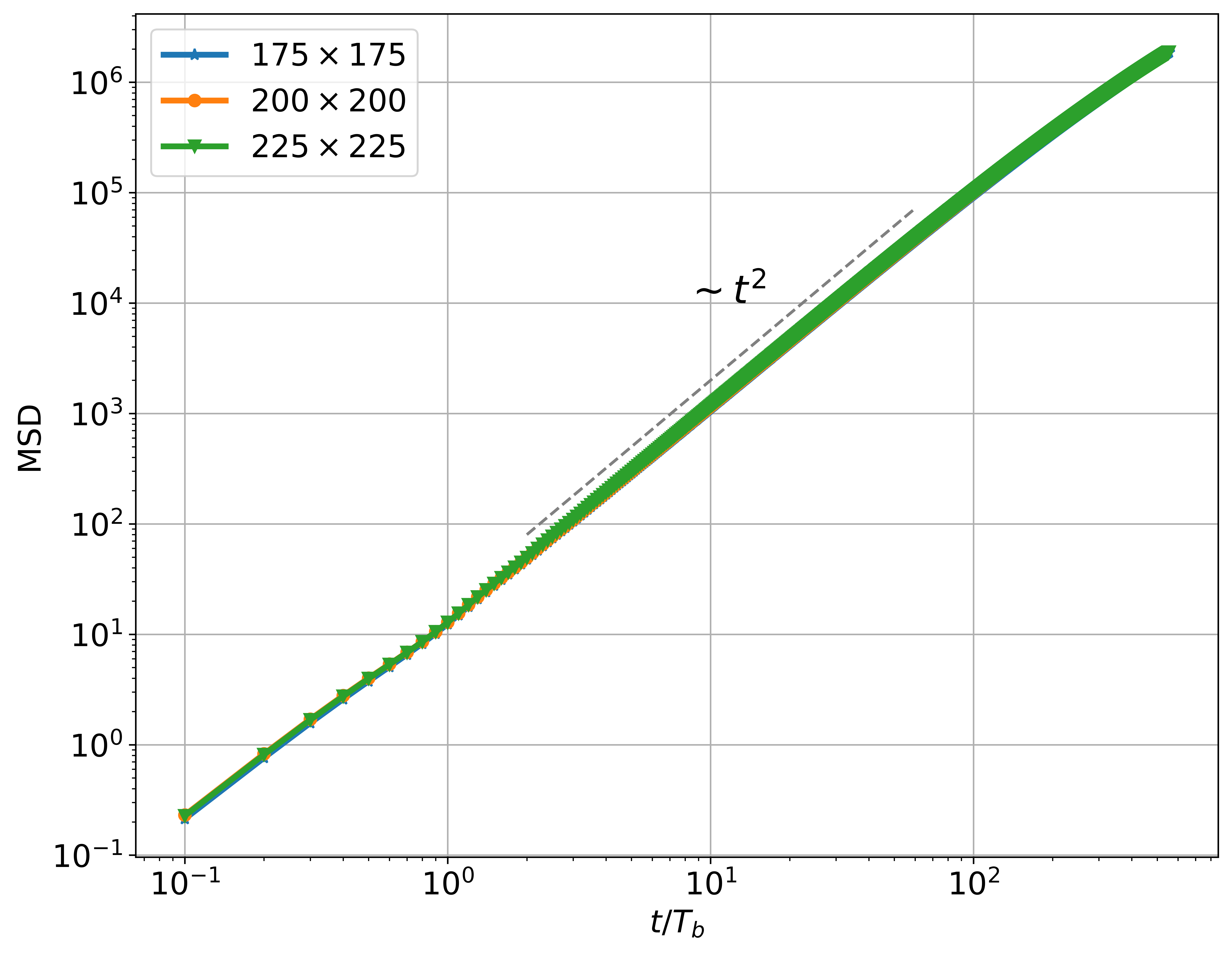}
    \caption{{\color{black}MSDs of $FA$ obtained for simulation boxes $175\times 175$, $200 \times 200$ and $225 \times 225$ respectively. All the MSDs fall on same curve indicating negligible finite size effects for simulation box of size $200 \times 200$.}}
    \label{fig:MSDFSE}
\end{figure}

\textcolor{black}{In systems exhibiting collective behavior, finite‐size effects emerge once the characteristic cluster size approaches the overall system size. As clusters form, their size grows over time according to a power law. Eventually, this length scale becomes comparable to the system’s dimensions, at which point no larger clusters can form. Consequently, the original power‐law scaling breaks down at long times, altering the asymptotic dynamics~\cite{krapivsky2010kinetic,venkata2021power}.}
\begin{figure}[!htbp]
    \centering
    \includegraphics[width=0.8\linewidth]{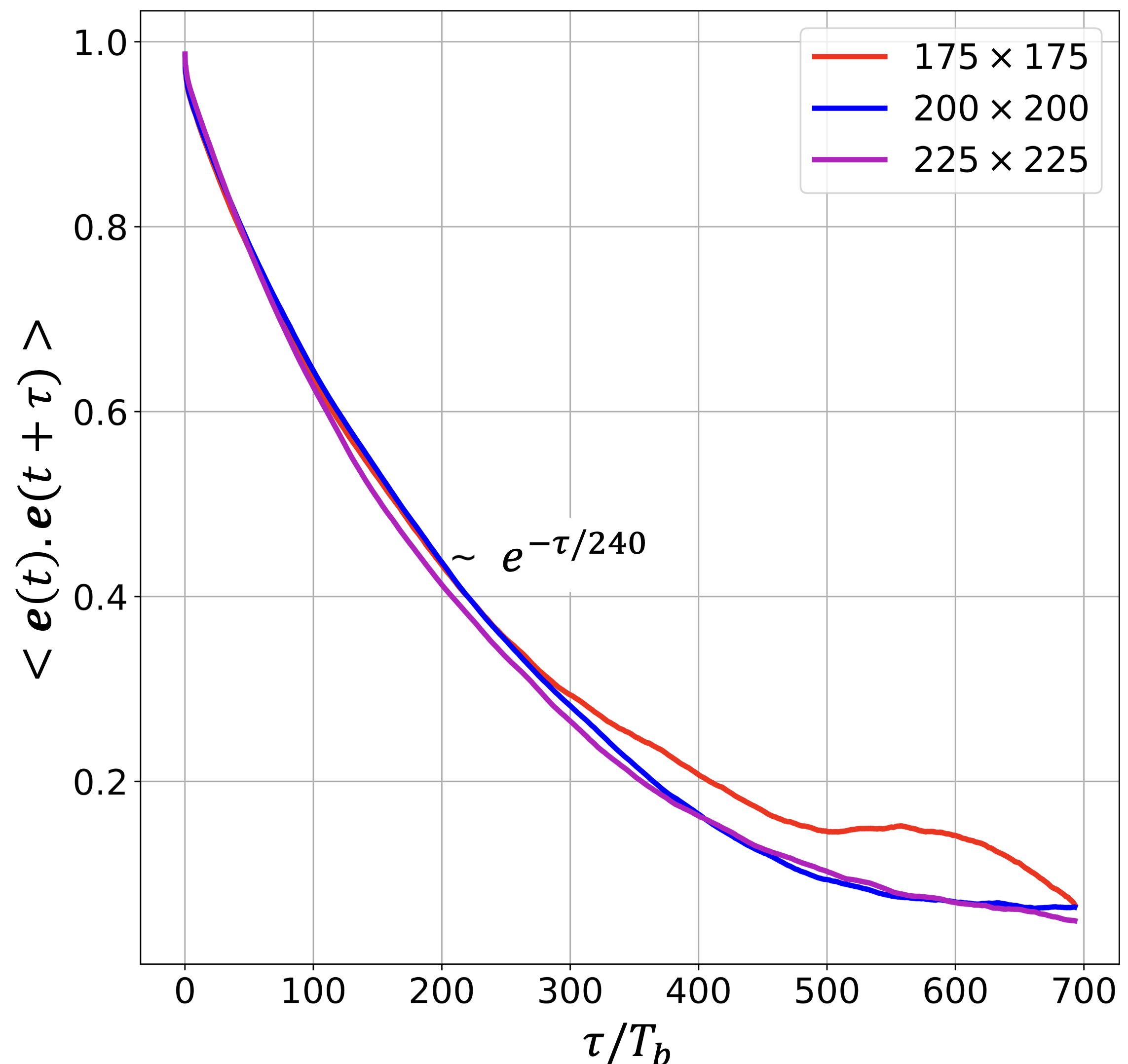}
    \caption{{\color{black}OCFs of $FA$ obtained for simulation boxes $175\times 175$, $200 \times 200$ and $225 \times 225$ respectively. OCFs from simulation boxes of sizes $200 \times 200$ and $225 \times 225$ closely match indicating negligible finite size effects for simulation box of size $200 \times 200$.}}
    \label{fig:OCFFSE}
\end{figure}

We use a \(200\times200\) simulation box, which is much larger than the size of an \(FA\). Yang \emph{et al.} demonstrated that finite‐size effects on clustering become negligible for boxes of size \(200\times200\) or larger~\cite{yang2008cooperation,yang2010swarm}. Because our study focuses on the dynamics of a single \(FA\) or a pair of \(FA\)s, we do not expect significant finite‐size effects. To confirm this, we computed the MSD and OCF of an \(FA\) in boxes of size \(175\times175\), \(200\times200\), and \(225\times225\). As shown in Figs.~\ref{fig:MSDFSE} and \ref{fig:OCFFSE}, all MSD curves overlap, and the OCFs for \(200\times200\) and \(225\times225\) coincide closely, indicating that finite‐size effects are negligible for our simulations.
\subsection{Ergodicity} 
In a non-ergodic process, time averages depend on the specific trajectory and therefore vary between realizations, rather than converging to a single ensemble average. Therefore, it is essential to examine the ergodicity of the process to determine whether time averaging yields reliable results.  

To quantify ergodicity breaking, Burov \emph{et al.}~\cite{burov2011single} defined the ergodicity breaking (EB) parameter as: 
\begin{equation}
	EB(t,T) = \frac{\left\langle \left(\overline{\Delta\bm x(t,T)\cdot \Delta\bm x(t,T)}\right)^2 \right \rangle}{\left \langle \overline{\Delta\bm x(t,T)\cdot \Delta\bm x(t,T)} \right \rangle^2}-1,
\end{equation}
where \( T \) represents the total simulation time.  For Brownian motion, in the limit \( t \ll T \), it has been shown that~\cite{metzler2014anomalous,deng2009ergodic,safdari2015quantifying,schwarzl2017quantifying,cherstvy2018time}:
\begin{equation}
	EB(t,T) \approx \frac{t}{T}.
\end{equation}

Figure~\ref{fig:EB} presents the EB parameter as a function of time lag for both active and inactive \( FA \). The Brownian motion limit of \( EB \) is also shown for reference. Clearly, the \( EB \) values for all \( FA \)s remain below the Brownian limit. This indicates that the dynamics of all \( FA \)s can be considered ergodic, meaning that time averages are representative of ensemble averages.  
\begin{figure}[t]
	\centering
	\includegraphics[width=0.7\columnwidth]{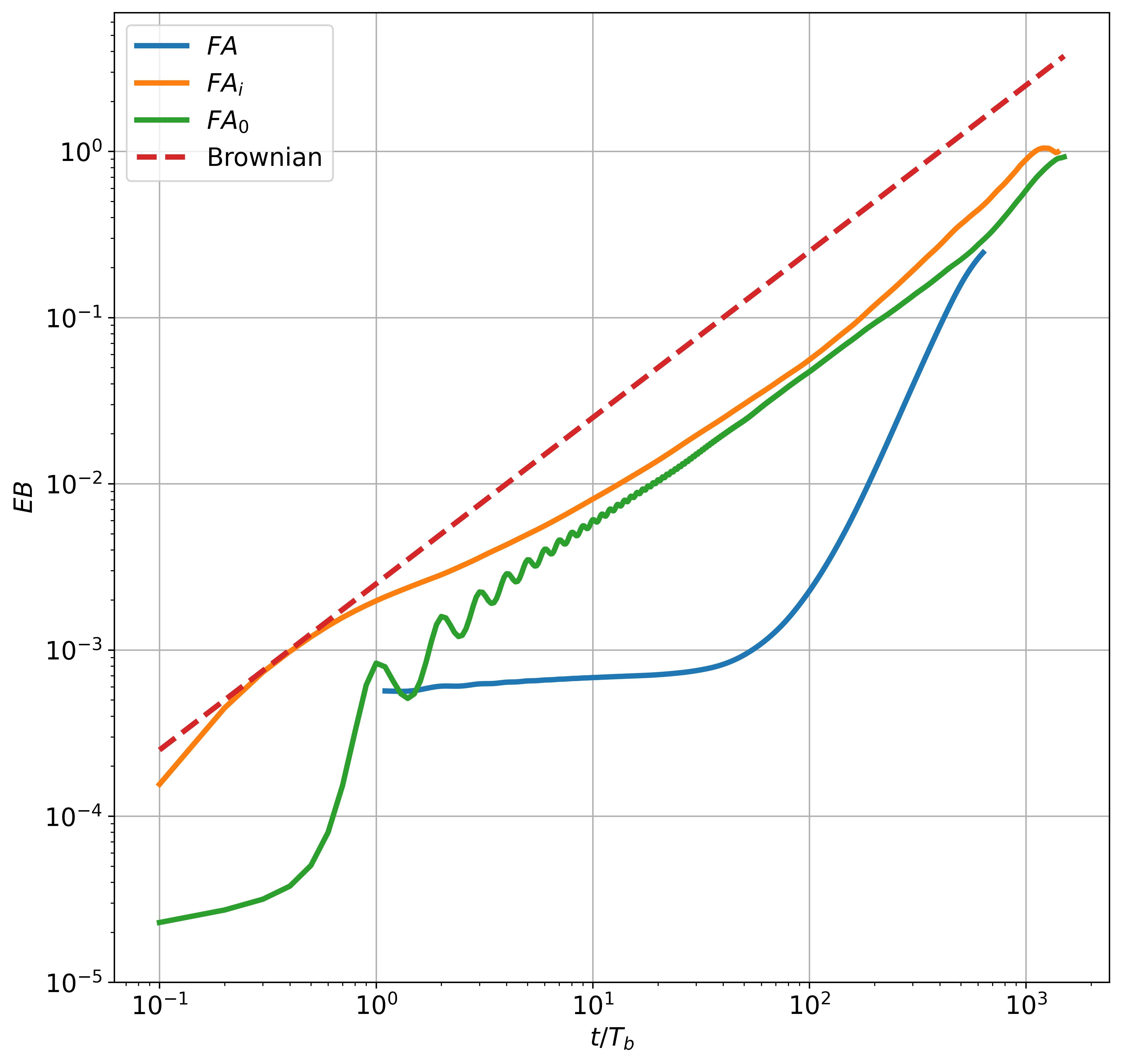}
	\caption{Ergodicity breaking (EB) parameter as a function of time lag for \( FA \), \( FA_0 \), \( FA_i \), and the approximate limit for Brownian motion. All the curves remaining below the Brownian limit indicate that the system is ergodic.}
	\label{fig:EB}
\end{figure}
}
\subsection{Supplemental Figures}
\begin{figure}[htbp!]
	\centering
	\includegraphics[width=0.8\columnwidth]{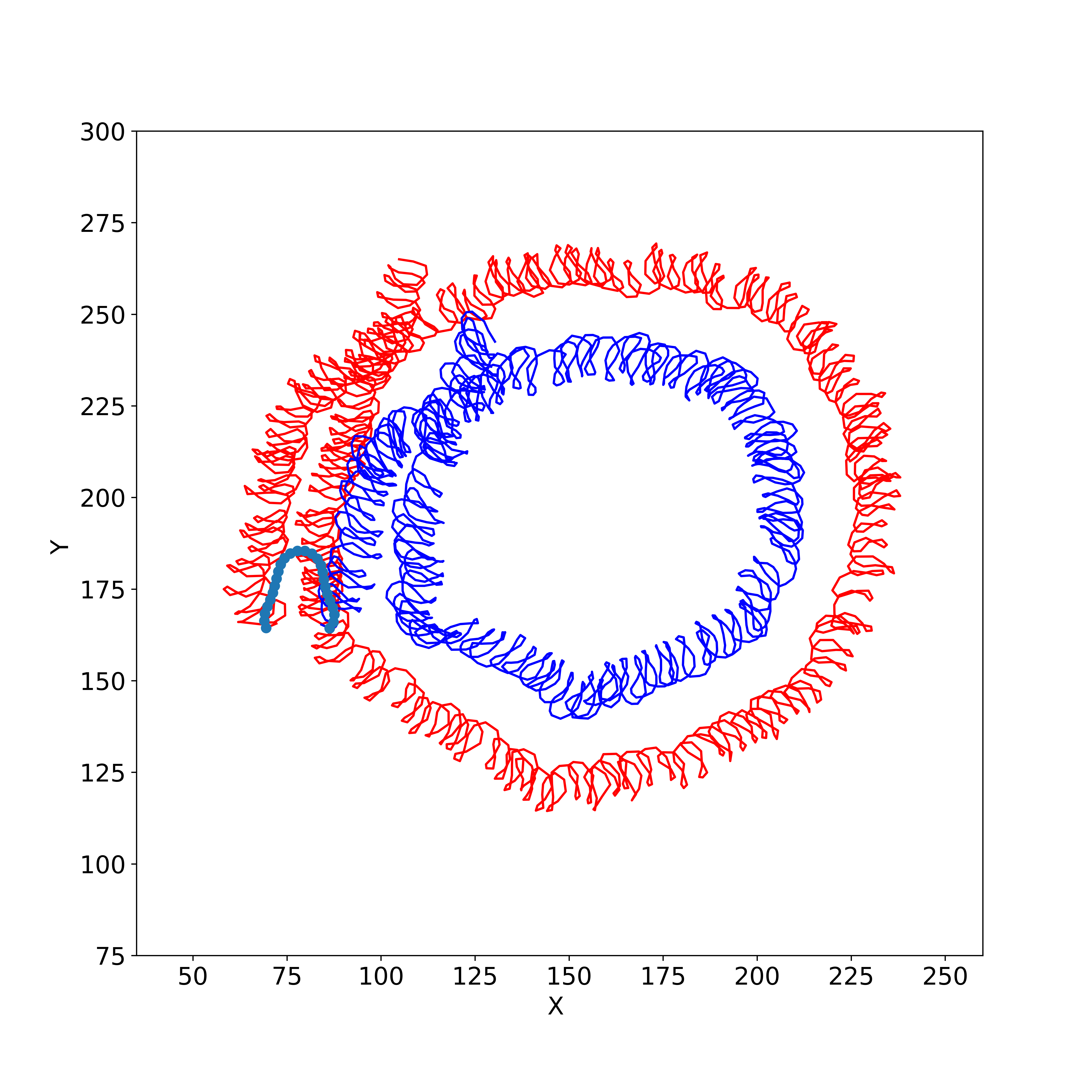}
	\caption{{\color{black}Trajectory of the endpoints of an \( FA \) with arms beating at frequencies \( f_1 = 1/120\) and \( f_2 = 0.7 f_1 \) (in units of \( \tau_{\text{MPC}}^{-1} \)). The arm with the lower frequency traces the inner blue trajectory. The \( FA \) is depicted at the initial instant for visual reference. The shape of the \( FA \) at the final time point is also shown; see also Video 7.}}
	\label{fig:Helical}
\end{figure}
\end{document}